\documentclass[twocolumn]{aastex631}

\usepackage{verbatim}
\usepackage{aas_macros}
\usepackage{hyperref}
\usepackage{natbib}
\usepackage{graphicx}
\usepackage[caption=false]{subfig}
\usepackage{epsfig}
\usepackage{amsmath}
\usepackage[english]{babel}

\newcommand\msun{M$_{\odot}$}
\newcommand\mstar{M$_{\rm{star}}$}
\newcommand\mbh{M$_{\rm{BH}}$}
\newcommand\Lbol{L$_{\rm{bol}}$}
\newcommand\lambdasBHAR{$\lambda_{\rm{sBHAR}}$}

\def\appropto{%
  \def\p{%
    \setbox0=\vbox{\hbox{$\propto$}}%
    \ht0=0.8ex \box0 }%
  \def\s{%
    \vbox{\hbox{$\sim$}}%
  }%
  \mathrel{\raisebox{0.7ex}{%
      \mbox{$\underset{\s}{\p}$}%
    }}%
}

\begin{document}


\title{Diverse pathways for supermassive black hole-galaxy coevolution}
\shorttitle{Diverse pathways for SMBH-galaxy coevolution}

\author[0000-0001-5529-7305]{Bryan A. Terrazas}
\affiliation{Department of Physics \& Astronomy, Oberlin College, Oberlin, OH, 44074, USA}

\author[0000-0003-1908-8463]{James Aird}
\affiliation{Institute for Astronomy, University of Edinburgh, Royal Observatory, Edinburgh EH9 3HJ, UK}

\author[0000-0002-2583-5894]{Alison L. Coil}
\affiliation{Department of Astronomy and Astrophysics, University of California, San Diego, 9500 Gilman Dr., La Jolla, CA 92093, USA}\

\correspondingauthor{Bryan Terrazas}
\email{bterraza@oberlin.edu}

\submitjournal{ApJ}

\begin{abstract}
Supermassive black holes (SMBHs) are observed in diverse galaxy populations across time yet a clear understanding of how they coevolve with their hosts has not been reached. Physically-motivated models of SMBH accretion and feedback vary widely between galaxy formation simulations due to the difficulty of modeling the range of scales important for galactic and SMBH processes. Here we use observational data to build an empirical model for SMBH growth. We apply observed specific accretion rate probability distributions as a function of galaxy star formation rate between $z = 0-2$ to the UniverseMachine galaxy formation model to determine SMBH accretion rates based on galaxy properties. We use observed $z = 0$ SMBH-stellar mass relations for the quiescent and star-forming populations to provide the local boundary conditions for SMBH growth histories. We then track the coevolutionary histories of galaxy stellar mass and their SMBHs backwards in time to $z = 2$. We find that the most massive SMBHs at $z = 0$ have grown very little of their total mass between $z = 0-2$, indicating early SMBH mass assembly for these systems. Conversely, lower mass SMBHs at $z = 0$ assembled their mass gradually across $z = 0-2$. This results in substantial evolution of the SMBH-stellar mass relation, shifting to higher normalization and shallower slope with increasing redshift. We find that the substantial scatter observed in the $z = 0$ SMBH-stellar mass relation results in the diversity of growth pathways found in our model, with some galaxies assembling their stellar mass before their SMBHs and others doing the opposite.
\end{abstract}



\keywords{galaxies: general -- galaxies: evolution -- galaxies: star formation}

\section{Introduction}
\label{sec:Intro}

There are compelling reasons for why supermassive black holes (SMBHs) are thought to play a crucial role in shaping galaxy evolution. Many of the most luminous and energetic events in the universe are produced by accretion onto these compact objects \citep{s1964, l1969, r1984}. For example, a Sagittarius A$^{*}$-like SMBH with a mass of $4\times10^{6}$ \msun\ accreting at its theoretical maximum, commonly called the Eddington limit, would release the equivalent of the Milky Way halo's binding energy in only $\sim$30 million years. This demonstrates the enormous potential for SMBHs to affect their host galaxies' evolution via feedback and accretion processes. \citep{f2012, hct2018}. This energetic output is thought to couple to the evolutionary processes of galaxies via preventative feedback, where it suppresses large-scale gas cooling from the circumgalactic medium, or ejective feedback, where it removes cold gas from the galaxy's interstellar medium \citep[ISM,][]{sd2015}. Some combination of both of these feedback mechanisms throughout the lifetime of a subset of galaxies eventually produces systems devoid of fuel for star formation, resulting in a quiescent galaxy population \citep{m2017, bcs2018}.

The accretion of material onto SMBHs produces light across the electromagnetic spectrum from X-ray to radio. The ability to identify active galactic nuclei (AGN) in galaxies by tracing the emission at particular wavelength regimes depends strongly on the viewing angle, the amount of obscuring material, and the physical state of the disk \citep{a1993, kht2003}. Star-forming galaxies with large cold gas reservoirs that can feed SMBH accretion are also subject to the most obscuration from their ISM \citep{fcf2018}. X-ray emission from the accretion disk is often the most reliable AGN detection method since it can penetrate high column densities \citep[with the exception of Compton-thick regions with equivalent hydrogen column densities $> 10^{24}$ cm$^{-2}$, e.g.][]{dcs2008} and provides a robust tracer of SMBH accretion across cosmic time.

Several studies have used the power of X-ray surveys to track SMBH growth over the evolving galaxy population. These studies find a broad distribution of X-ray AGN luminosities within galaxies of a given star formation rate \citep[SFR,][]{aca2017}. While star-forming galaxies are more likely to host an AGN \citep{srs2012}, the X-ray luminosity of these galaxies has a very broad distrubiton that is only weakly dependent on their host galaxy SFR \citep{acm2012, ham2012, aac2015, rsm2019}. The broad distribution of luminosities likely reflects variability in the accretion rate of SMBHs at much shorter timescales compared to changes in global galactic properties such as SFR \citep{hma2014}. Once the variability on galactic timescales is accounted for, a complex but more direct connection between AGN feeding and star formation is revealed \citep{acg2019,ack2022}.

In the local Universe, we find the end result of this accretion-powered growth: a diverse galaxy population hosting SMBHs with a wide range of masses identified using dynamical tracers at $z\sim0$ \citep{kh2013,gsh2020}. While SMBH mass (\mbh) broadly correlates with galaxy properties, there is substantial scatter in \mbh\ at a given host galaxy stellar mass (\mstar). SMBHs that are overmassive relative to their host's \mstar\ tend to be found in galaxies with lower star formation activity \citep{tbh2016, tbw2017, mbr2018, pbm2022}. SMBH mass is therefore an important indicator of quiescence in galaxies, at least in the local Universe where high-fidelity dynamical SMBH masses can be measured \citep{kh2013, soe2016, v2016}. 

However, these \mbh\ measurements are difficult to obtain for a representative sample of galaxies since it is necessary to resolve and detect unobscured dynamical tracers within the SMBH's gravitational sphere of influence, introducing selection effects that are poorly understood \citep{sbs2016, sbr2019}. Using a variety of methods to trace the dynamical mass (e.g., stars, gas, reverberation mapping) allows for SMBH masses to be measured in several types of host galaxy, somewhat alleviating this problem. At higher redshifts, resolving the sphere of influence is impossible with current observational capabilities. While some studies have attempted to place constraints on the evolution of SMBH scaling relations for AGN out to $z \sim 2$ \citep{sct2020, m2023, lsh2023, tsd2024}, the samples used are heavily biased towards the most massive and highly accreting SMBHs. As such, determining the representative coevolutionary pathway(s) taken by SMBHs and their host galaxies to reproduce the observed $z = 0$ relations remains an open question.

Galaxy formation simulations have historically relied on SMBH feedback for suppressing star formation in massive galaxies and thus reproducing the observed stellar mass functions and quenched fractions across redshift \citep{dsh2005, bbm2006, cfb2009}. More massive SMBHs are observed to live in more massive galaxies \citep{fm2000, gbb2000}, and simulations have broadly reproduced the importance of SMBH mass as an indicator for quiescence. However, physical models of SMBH growth and feedback still vary substantially between the latest state-of-the-art simulations \citep{tbh2016, tbw2017, hls2021, dnp2024}.

Hydrodynamical simulations of galaxy formation that model regions of space $\gtrsim$ 1 Mpc lack the resolution needed to capture the processes governing SMBH physics at scales of $\ll$ 0.1 pc. This reflects the difficult task of simulating the large range of spatial and temporal scales necessary for modeling galactic processes, particularly those pertaining to the central SMBH. In order to overcome this challenge simulators often employ sub-grid physics that model the effects of sub-resolution scales on the smallest resolution elements in the simulations \citep[e.g.,][]{csw2006, hds2014, svg2015, csb2015, wsp2018}. These sub-grid models for SMBH accretion and feedback differ greatly between galaxy formation simulations due to the lack of a consensus model for these processes, motivating the need to further constrain these models with observational data. Despite the differences in SMBH subgrid models, different simulations successfully reproduce the suppression of star formation in massive galaxies at late times \citep{tbh2016}. This indicates that other metrics that are more sensitive to the differences between SMBH models are necessary to fully understand how a quiescent population is produced \citep{spd2022}. One such powerful metric that reflects the differences in simulations is the relationship between SMBH mass and total host galaxy stellar mass \citep{tbh2016, tbw2017}. In this work, we focus on better understanding how this relation evolves over time.

Here, we use observational data to build an \textit{empirical} model of SMBH growth. These data place constraints on SMBH masses at $z = 0$ and accretion rates at $z = 0-2$, as a function of host galaxy star formation rate (SFR). Our model uses the observed \mbh\---\mstar\ relations for quiescent and star-forming galaxies at $z = 0$ from \citet{gsh2020} as the local boundary conditions. We also use the observed probability distribution functions for specific SMBH accretion rate (sBHAR) as a function of both redshift and SFR from \citet{acg2019}, who carefully correct their distributions for incompleteness using a Bayesian approach. Our goal is to use these data to understand how the $z = 0$ distribution of galaxies on the \mbh\---\mstar\ relation established itself from $z = 2$ to the present day. This exercise will provide constraints on how SMBH growth proceeds within a diversity of galaxies based on observational data.

We begin with a description of the observational data used in our empirical approach and an explanation of our model in Section~\ref{sec:empmodel}. We then describe the evolution of the \mbh\---\mstar\ relation from $z = 0-2$ in Section~\ref{sec:scalingrelations}, indicating the diversity of possible coevolutionary growth histories. Section~\ref{sec:phase} describes how much SMBH growth occurs in different phases of star formation across time. Section~\ref{sec:highMbh} focuses on the small fractional mass growth since $z = 2$ of the most massive SMBHs at $z = 0$, indicating the early assembly of these objects in a subset of massive galaxies. We then describe the impact of varying the $z = 0$ \mbh\---\mstar\ relation which serve as the local boundary conditions for our model in Section~\ref{sec:varrel}. We end with a discussion of our most interesting results in Section~\ref{sec:discussion} and a summarizing conclusion in Section~\ref{sec:conclusions}. Throughout this paper, we adopt a flat cosmology with $\Omega_{\Lambda} = 0.7$ and $H_0 = 70$ km s$^{-1}$ Mpc$^{-1}$.





\section{An empirical model for SMBH growth in galaxies}
\label{sec:empmodel}

Developing an {\it a priori} physical model of SMBH growth within a cosmological context has proven difficult due to the large range of scales important for SMBH accretion and galactic evolution. We aim to provide a complementary approach to study these growth histories by building an empirical model of SMBH growth within a galaxy formation model. In this work we use observational constraints to explore how SMBH masses are distributed amongst galaxies of different stellar masses and SFRs at earlier epochs. This section describes the empirical galaxy formation model we build these SMBH growth histories upon, the boundary conditions for these growth histories at $z = 0$, and the observational constraints used to allocate growth rates to SMBHs in the simulated galaxies as a function of their SFR across time. Finally we describe how we build the model, what assumptions are made, and a few model variations that alter these assumptions.

\subsection{The UniverseMachine empirical model}
\label{sec:um}

The UniverseMachine is an empirical model that builds galaxy growth histories using observational constraints \citep{bwh2019}. The model builds the baryonic component of galaxies on top of a dark matter-only simulation, in this case the $\it{Bolshoi}$-$\it{Planck}$ simulation\footnote{The UniverseMachine has also been run on the \textit{MDPL2} dark matter-only simulation which has a larger volume and lower resolution, which we do not use for our work.} with a comoving volume of 250 $h^{-1}$ Mpc$^{3}$ \citep{ktp2011, rbp2016}. SFRs are assigned to halos based on the halo potential well depth, halo assembly history, and redshift of a given halo. From these, stellar masses and UV luminosities are determined and galaxy catalogs are created corresponding to a range of evenly spaced snapshots ($\Delta t \sim 0.1$ Gyr) across cosmic time. These galaxy catalogs are iteratively compared to observational constraints in a Markov Chain Monte Carlo algorithm and the parametrizations used to assign SFRs are adjusted until they produce agreement with the observations. The observational constraints used to determine agreement with the real Universe are stellar mass functions, UV luminosity functions, the UV-stellar mass relation, specific and cosmic SFRs, central galaxy quenched fractions as a function of environmental density, and galaxy autocorrelation functions. These observations span $0 < z \lesssim 10.5$, depending on the measurement, and thus enable an empirical characterisation of galaxy evolution over a broad swathe of cosmic history.

We note that the UniverseMachine incorporates the effects of stellar mass loss due to winds and stellar recycling into their framework. This can result in stellar mass growth histories that show negative growth with time for some quiescent galaxies, as seen in some of the figures in this paper.

In this work we add SMBHs into the framework of the UniverseMachine in post-processing, expanding on the approach taken by \citet{ac2021}. Just as the UniverseMachine uses observational constraints with limited physical priors, we too build an empirical model that reproduces the observational results described in the next two sections without assuming the physical processes governing SMBH accretion. Since SMBH-galaxy correlations are best known locally, we begin by distributing SMBHs of different masses across the UniverseMachine galaxy population at $z = 0$ according to the observations. We then build growth histories backwards in time by assigning SMBH accretion rates depending on the galaxy SFR and redshift according to other observations. This procedure is described further in Section~\ref{sec:method} and shown schematically in Fig.~\ref{fig:method}.


\subsection{Probability distributions of SMBH accretion rates from z=0--3}
\label{sec:mdotbh}

\citet{acg2018, acg2019} determine probability distributions of specific SMBH accretion rates (sBHAR) as a function of host galaxy properties \citep[see also][]{acm2012, gas2017, ybv2018}. These works use a sample of galaxies from four of the CANDELS fields and the UltraVISTA survey. They extract X-ray data for each galaxy from Chandra observations in the same fields. They use a Bayesian approach to account for the sensitivity limits of Chandra and constrain probability distributions of SMBH accretion rates that are corrected for incompleteness in the X-ray detections. These works define the SMBH accretion rate in relation to their host galaxy stellar mass, resulting in a specific black hole accretion rate, \lambdasBHAR. This quantity is defined as

\begin{equation}
\lambda_{\rm{sBHAR}} = \frac{L_{\rm{bol}}}{1.3\times10^{38} \rm{erg}\; \rm{s}^{-1} \times 0.002 \dfrac{M_{\rm{star}}}{M_{\odot}}},
\label{eqn:lambda}
\end{equation}
where $L_{\rm{bol}} = k_{\rm{bol}} L_{\rm{X}}$. Here, $L_{\rm{X}}$ is the 2-10 keV X-ray luminosity and $k_{\rm{bol}}$ is the bolometric correction factor assumed to be constant at a value of 25 \citep{acg2018}.

The denominator in this equation was chosen by \citet{acg2018, acg2019} such that \lambdasBHAR\ $\approx L_\mathrm{bol}/L_\mathrm{Edd}$, the Eddington ratio ($f_\mathrm{Edd} = L_\mathrm{bol}/L_\mathrm{Edd}$) assuming a \textit{fixed} scaling between \mbh\ and \mstar. It is important to note that in our analysis we incorporate the observed intrinsic scatter between \mbh\ and \mstar\ and its correlation with SFR. This scatter propogates to higher redshift and a fixed and tight scaling between \mbh\ and \mstar\ is not assumed. Thus in our analysis \lambdasBHAR$\neq f_\mathrm{Edd}$ in general (see Fig.~\ref{fig:feddlambda}).

Each of these works splits the galaxy population differently, exploring the dependence of these sBHAR distributions as a function of (1) stellar mass for star-forming and quiescent galaxies separately \citep{acg2018} and (2) SFR relative to the star forming main sequence \citep{acg2019}. In this work we use the probability distributions of sBHAR described in \citet{acg2019} that selected galaxies over a broad stellar mass range ($10.0<\log ($\mstar/\msun$)<11.5$) and split this population into five bins relative to the star-forming main sequence: starburst, main sequence, sub-main sequence, high quiescent, and low quiescent. Probability distributions for sBHAR are determined for each of these SFR bins at different redshifts (see Section~\ref{sec:method}). Galaxies are binned into $z = 0.1-0.5$, $0.5-1.0$, $1.0-1.5$, and $1.5-2.0$ redshift ranges and measurements of the \lambdasBHAR\ probability distributions are performed independently for each of the SFR bins. We aim to explore how SMBHs grow alongside the galaxy population as galaxies move between these different SFR bins across cosmic time. We limit our exploration to galaxies with \mstar\ $> 10^{10}$ \msun\ and redshifts $0 < z < 2$ since the uncertainties on the observed accretion rates become large for galaxies at lower masses and higher redshifts.

\subsection{SMBH Masses at z=0}
\label{sec:mbh}

Many studies have found a relationship between dynamical SMBH mass and host galaxy properties \citep{fm2000, hr2004, grg2009, mm2013}. Increasing evidence indicates that quiescent galaxies host more massive SMBHs compared to star forming galaxies at the same stellar mass \citep{tbh2016, tbw2017, mbr2018, pbm2022}. \citet{tbh2016, tbw2017} used far-infrared data from \textit{IRAS} \citep{nhv1984} to estimate SFRs of galaxies with dynamically-measured SMBH masses and found that the star formation properties correlate closely with the ratio between SMBH mass and stellar mass at $z = 0$. Due to the difficulty in measuring extragalactic SMBH masses dynamically, current observational constraints limit studies like this to the local Universe. 

\begin{figure*}
\centering
\subfloat{\includegraphics[width = 5.6cm]{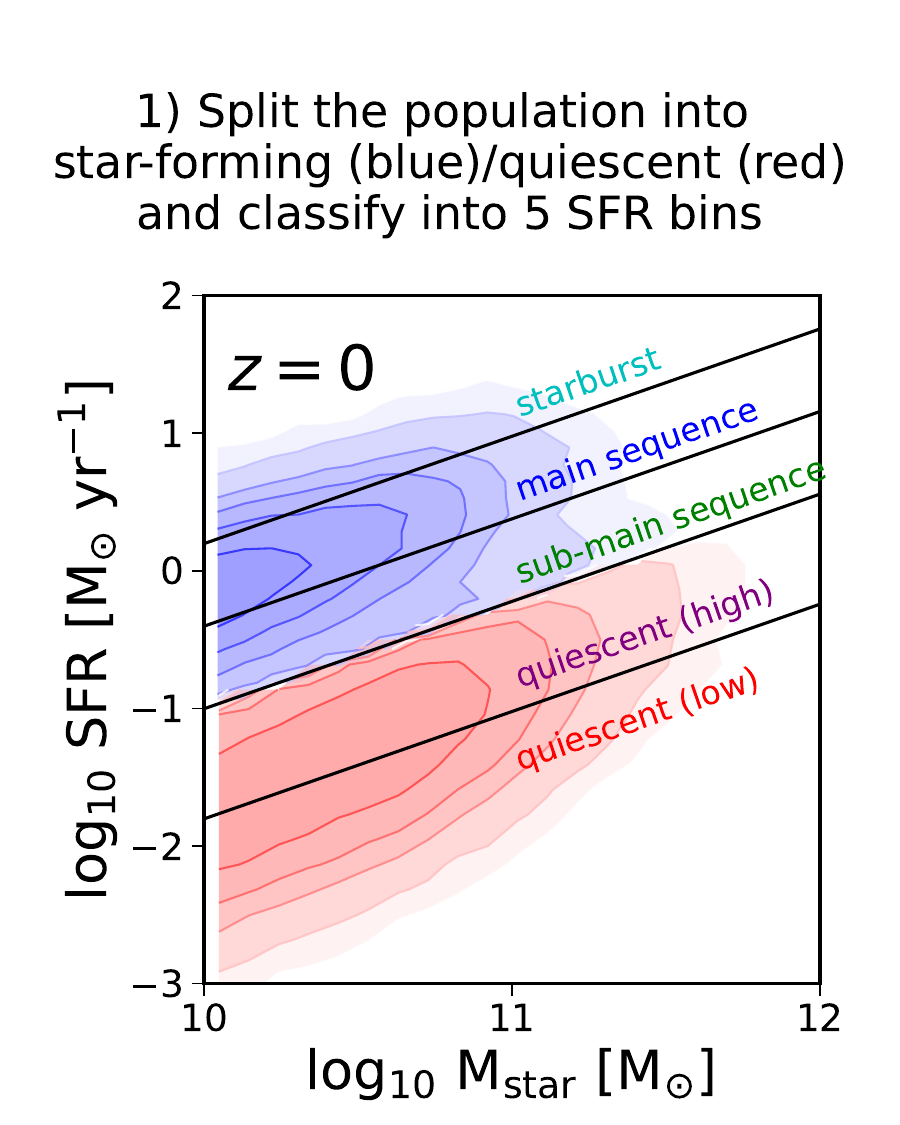}}
\subfloat{\includegraphics[width = 5.6cm]{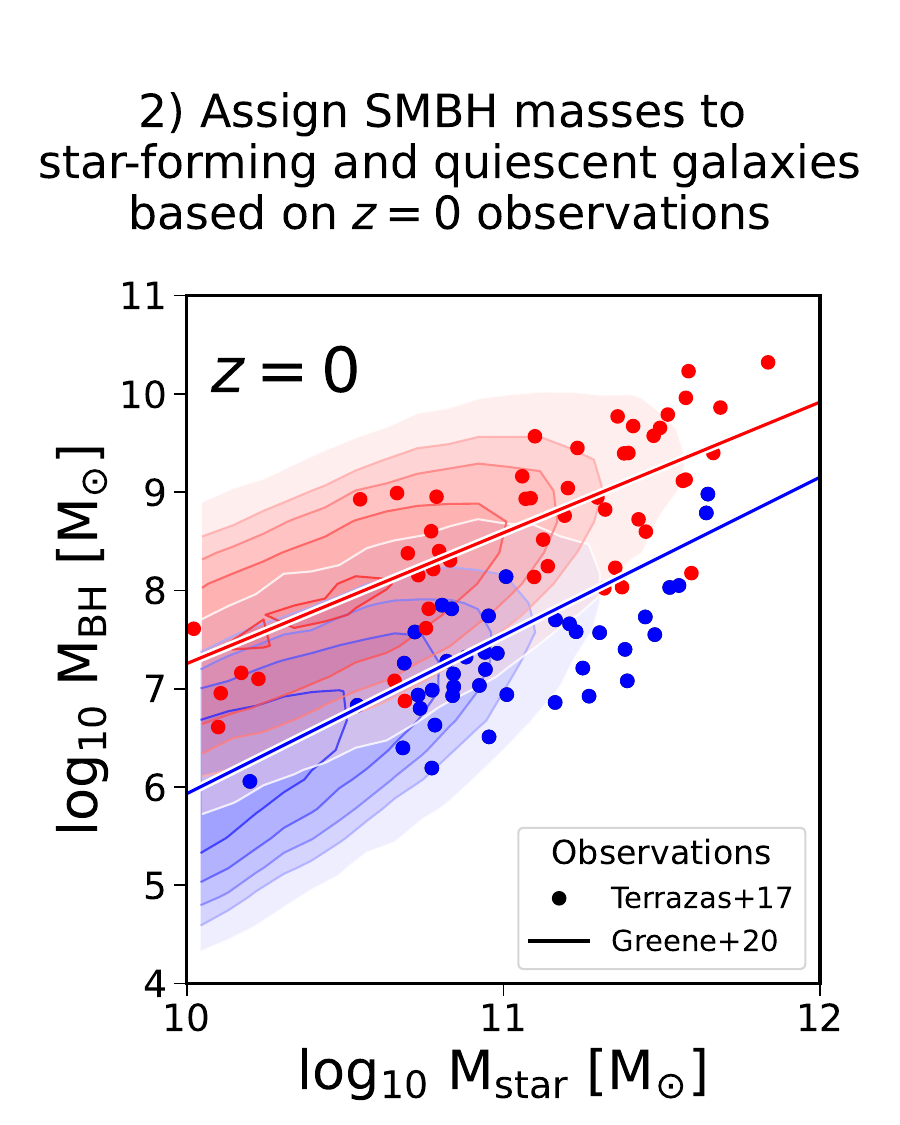}}\\
\subfloat{\includegraphics[width = 5.6cm]{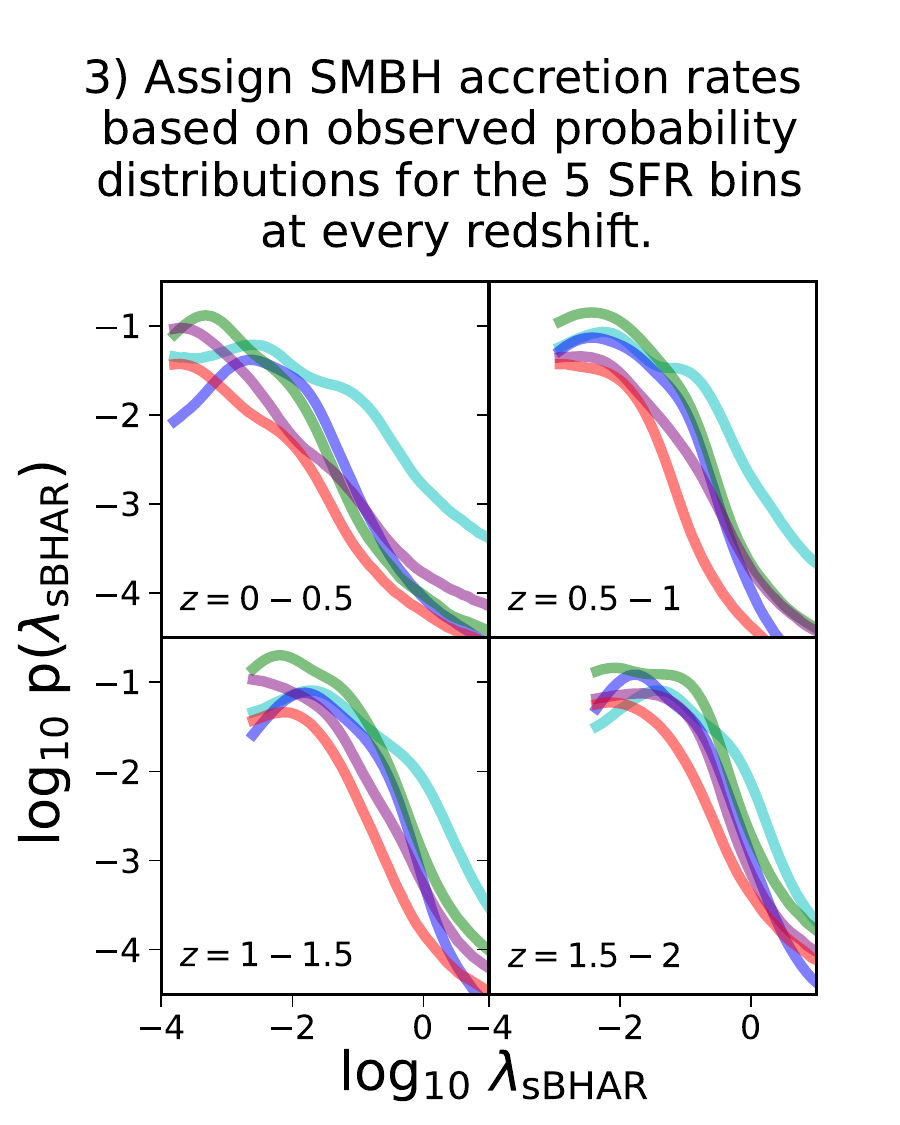}}
\subfloat{\includegraphics[width = 5.6cm]{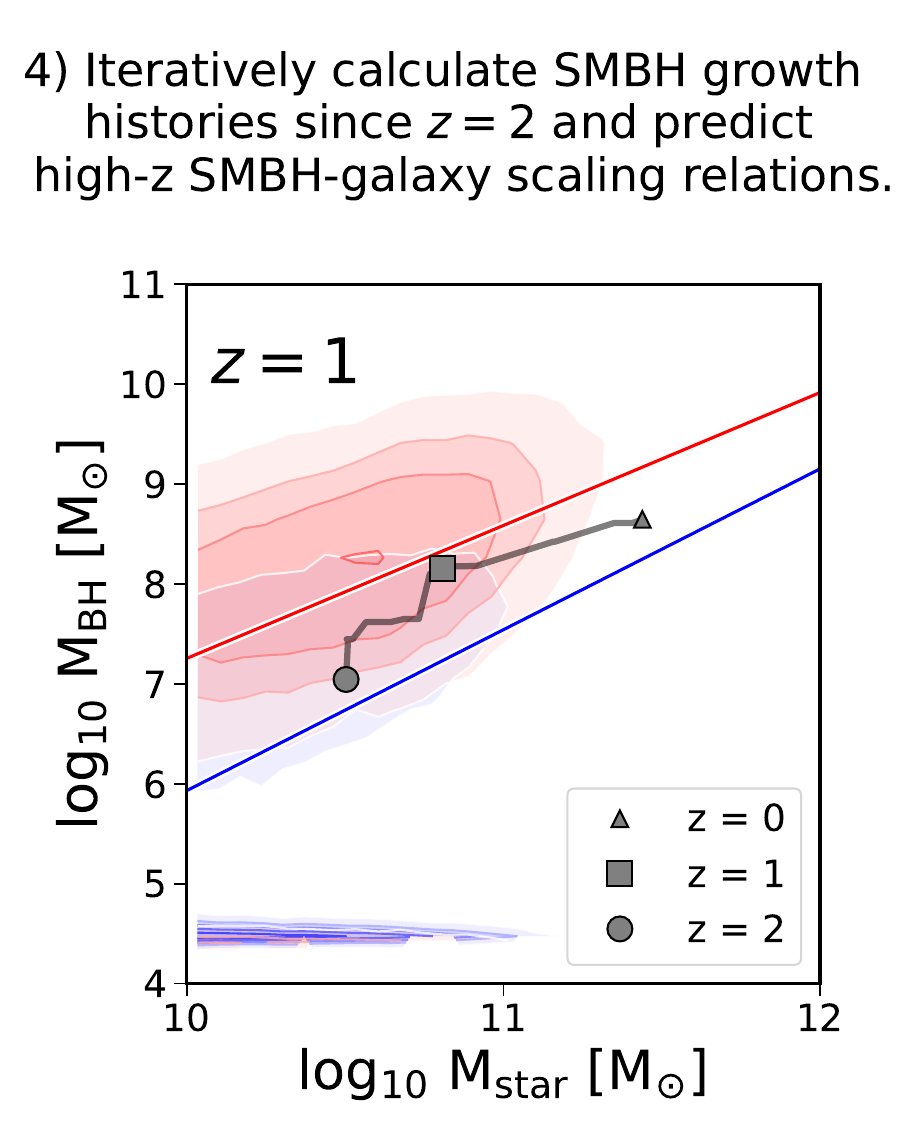}}
\caption{The procedure used for building empirically-motivated SMBH growth histories. We begin with $z = 0$ galaxies from the UniverseMachine empirical galaxy formation model \citep{bwh2019}, splitting them into five populations based on SFR relative to the main sequence (step 1). We then assign SMBH masses to simulated galaxies based on the observed SMBH mass-stellar mass scaling relations for quiescent and star-forming galaxies at $z = 0$ from \citet{gsh2020} (step 2). We then assign specific SMBH accretion rates (sBHAR) to simulated galaxies based on observationally-derived probability functions from \citet{acg2019} (step 3) and subtract the estimated amount of SMBH mass grown between two adjacent simulation timesteps. We re-assign sBHARs at every snapshot based on galaxy SFR and redshift. We then determine SMBH growth histories for each galaxy from $z = 2$ to $z = 0$ (step 4). The contours in this last panel indicate the $z = 1$ distribution of galaxies that are progenitors of the $z = 0$ star-forming (blue) and quiescent (red) populations. The red and blue lines show the $z = 0$ \citet{gsh2020} relations. The grey line shows an example of a galaxy coevolutionary growth track from $z = 0$ to 2 from our model.}
\label{fig:method}
\end{figure*}

\citet{gsh2020} provide \mbh\---\mstar\ relations defined separately for early- and late-type galaxies using an updated sample of dynamical SMBH masses. We use these relations to assign a $z = 0$ SMBH mass to galaxies in the UniverseMachine:
\begin{equation}
\rm{log}_{10} M_{\rm{BH}}(z = 0) = \alpha + \beta \rm{log}_{10} \dfrac{M_{\rm{star}}}{M_0} + \epsilon
\label{eqn:greene}
\end{equation}
where $\alpha = 6.70$, $\beta = 1.61$ for late-type galaxies and $\alpha = 7.89$, $\beta = 1.33$ for early-type galaxies. In this work late-type corresponds to star-forming galaxies and early-type to quiescent galaxies. For both populations $M_0 = 3\times10^{10}$ \msun\ and the intrinsic scatter is modeled as a Gaussian with a standard deviation $\epsilon = 0.65$. This produces a quiescent population with more massive SMBHs than the star-forming population for a given \mstar\ (see upper right panel of Fig.~\ref{fig:method}). These observationally-derived scaling relations provide the boundary conditions from which we build SMBH growth histories backwards in time.

In Section~\ref{sec:varrel} we introduce two alternative methods to assign SMBH masses to the $z = 0$ galaxy population. In one variation (Model Variation 3) we assume a strong correlation between SFR, \mstar, and \mbh\ at $z = 0$ using Eqn. 1 from \citet{tbw2017},
\begin{equation}
\begin{aligned}
(0.82\pm\epsilon_1)~\rm{log}_{10} M_{\rm{BH}}(z = 0) = \\
(0.8\pm\epsilon_2)~\rm{log}_{10}\dfrac{M_{\rm{star}}}{M_{\rm{star,0}}} \\ -~ \rm{log}_{10} \dfrac{SFR}{{M_{\rm{star}}}} \\-~ (11.84\pm\epsilon_3)
\end{aligned}
\label{eqn:t17}
\end{equation}
where the scatter in this three dimensional space is modeled as Gaussians with standard deviations $\epsilon_1 = 0.08$, $\epsilon_2 = 0.18$, and $\epsilon_3 = 0.10$. This model represents a more extreme dependence between SMBH mass and SFR than our fiducial approach, where the specific star formation rate (sSFR) is approximately inversely proportional to the SMBH-stellar mass ratio, sSFR $\propto \dfrac{M_{\rm{star}}^{0.8}}{M_{\rm{BH}}^{0.82}} \appropto \dfrac{M_{\rm{star}}}{M_{\rm{BH}}}$.

In another variation (Model Variation 4) we assume a strong correlation between \mbh\ and \mstar\ following \citet{kh2013},
\begin{equation}
M_{\rm{BH}}(z = 0) = 4.9\times10^{8} \left(\dfrac{M_{\rm{star}}}{10^{11} M_{\odot}}\right)^{1.16},
\label{eqn:kh13}
\end{equation}
with an intrinsic scatter of 0.29. This model represents the other extreme where there is no correlation with SFR. Galaxy formation models often rely on calibrating or comparing their models to correlations such as this one without taking into account the intrinsic scatter between \mbh\ and \mstar\ that correlates with SFR. The results of both of these model variations are discussed further in Section~\ref{sec:varrel}.

\subsection{Building SMBH growth histories}
\label{sec:method}

The previous three sections describe the UniverseMachine empirical model and the two sets of observational constraints that we use to build SMBH growth histories. In this section we describe our fiducial method for building SMBH growth histories onto the UniverseMachine in post-processing using these constraints. 

\subsubsection{Assigning SMBH masses}

The first step is to categorize galaxies in the UniverseMachine as star-forming or quiescent at $z = 0$ using a SFR cut, which we define as being 0.9 dex below the star-forming main sequence (SFMS). For UniverseMachine galaxies we find a fit to the SFMS with the following equation:
\begin{equation}
\rm{log}_{10} SFR_{\rm{MS}} = -7.9 + 0.78 log_{10} M_{\rm{star}} + 3 \rm{log}_{10}(1+z)
\label{eqn:SFMS}
\end{equation}
We then assign SMBH masses to $z = 0$ galaxies differently depending on whether they are star-forming or quiescent following the observed $z = 0$ \mbh\---\mstar\ relations for these two populations from \citet{gsh2020} (see Section~\ref{sec:mbh}).

\subsubsection{Assigning sBHARs}
\label{sec:assign}

The next step is to assign \lambdasBHAR\ values to simulated galaxies by randomly sampling from the sBHAR probability distributions provided by \citet{acg2019} for that galaxy's SFR classification. 

We define these galaxy classifications with the following SFR bins:
\begin{itemize}
  \item starburst: log$_{10}$ $\dfrac{\rm{SFR}}{\rm{SFR}_{\rm{MS}}}$ $>$ 0.3
  \item main sequence: -0.3 $<$ log$_{10}$ $\dfrac{\rm{SFR}}{\rm{SFR}_{\rm{MS}}}$ $<$ 0.3
  \item sub-main sequence: -0.9 $<$ log$_{10}$ $\dfrac{\rm{SFR}}{\rm{SFR}_{\rm{MS}}}$ $<$ -0.3
  \item high quiescent: -1.7 $<$ log$_{10}$ $\dfrac{\rm{SFR}}{\rm{SFR}_{\rm{MS}}}$ $<$ -0.9
  \item low quiescent: log$_{10}$ $\dfrac{\rm{SFR}}{\rm{SFR}_{\rm{MS}}}$ $<$ -1.7
\end{itemize}

These SFR bin classifications and the fit to the star forming main sequence (Eqn.~\ref{eqn:SFMS}) differ from the definitions described in \citet{acg2019} because the observational CANDELS/UltraVISTA data used in that paper and the results from the UniverseMachine produce quantitatively different distributions on the SFR-\mstar\ relation. The SFR bins we define (see upper left panel of Fig.~\ref{fig:method}) produce a qualitatively better match to the categories presented in \citet{acg2019} at all redshifts we explore. 

Alongside this adjustment to the binning and SFMS, all UniverseMachine galaxies with SFRs that lie more than 1 dex below SFR$_{\rm{SFMS}}$(z) are adjusted to higher values by adding SFR$_{\rm{SFMS}}$(z)-SFR$_{\rm{SFMS}}$($z=0$) to the value given by UniverseMachine. This produces a quiescent population that lies a consistent distance from the evolving SFMS, resulting in a better agreement between the observed SFR distributions of high redshift galaxy populations and the UniverseMachine results. This adjustment is required for the boundary between our classifications of high and low quiescent galaxies to evenly split the quiescent population between these two bins as is described in \citet{acg2019}. We confirm that this SFR adjustment would add a negligible amount of mass growth throughout these galaxy's histories and primarily serves the purpose of more easily splitting the galaxy population into SFR bins.

According to this procedure all starburst, main sequence, and sub-main sequence galaxies are defined as star-forming, while the other two bins are quiescent (upper left panel of Fig.~\ref{fig:method}). The probability distributions in \citet{acg2019} (lower left panel of Fig.~\ref{fig:method}) are integrated to determine the X-ray-bright AGN fraction at each redshift:
\begin{equation}
\rm{f}_{\rm{AGN,X}} = \int_{-4}^{\infty} p(\lambda_{\rm{sBHAR}})\; \rm{d\;log}\lambda_{\rm{sBHAR}}
\label{eqn:fagn}
\end{equation}
where p(\lambdasBHAR) is the probability distribution for each SFR classification bin. We take into account obscured AGN that are not visible in the X-rays by assuming $\rm{f}_{\rm{AGN}} = 2 \times \rm{f}_{\rm{AGN,X}}$. For each UniverseMachine snapshot, we choose a random subset of galaxies of size $N \times f_{\rm{AGN}}$, where $N$ is the total number of galaxies in that SFR bin and assign \lambdasBHAR\ values to only those galaxies. We  sample only \lambdasBHAR\ values $< 1.0$ at $z < 1.0$, while at $z > 1.0$ \lambdasBHAR\ values may reach values up to 10.0.

\citet{acg2019} provide probability distributions for 5 SFR classification bins at each of the following redshift bins: $z = 0.1-0.5$, $0.5-1.0$, $1.0-1.5$, $1.5-2.0$, and $2.0-2.5$ (see lower left panel of Fig.~\ref{fig:method}). In our model, we assume each of these curves represents the center of that redshift bin ($z = 0.3, 0.75, 1.25, 1.75,$ and $2.25$). We use linear interpolation to evolve these probability distributions between these redshift bins and linear extrapolation to evolve the relation between $z = 0.3$ to $0$. This is necessary to avoid discontinuities in the resulting SMBH growth histories had we assumed the probability distributions change abruptly between redshift bins.

In Section~\ref{sec:highMbh} we vary the assumption that the \lambdasBHAR\ distribution is randomly sampled. In one variation (Model Variation 1) we rank order galaxies based on their \mbh/\mstar\ ratio, preferentially assigning higher \lambdasBHAR\ values to those with overmassive SMBHs. In another (Model Variation 2) we only assign \lambdasBHAR\ values to galaxies with high \mbh/\mstar\ ratios (see Fig.~\ref{fig:mvs}). Both of these adjustments serve to preferentially place SMBH growth within the most massive SMBHs in the universe. We explore these variations for assigning \lambdasBHAR\ within each SFR classification in Section~\ref{sec:highMbh}.

\begin{figure}
\centering
\epsfxsize=8.5cm
\epsfbox{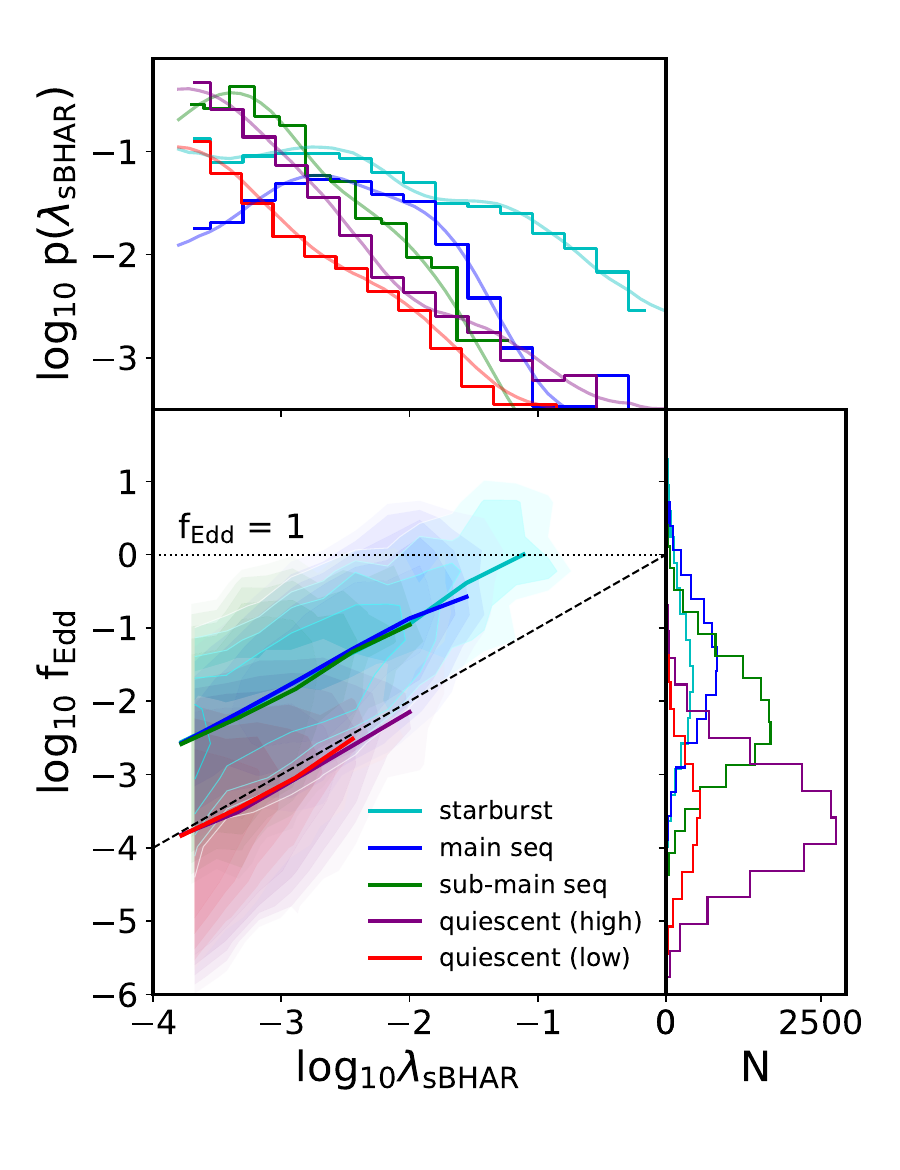}
\caption{The Eddington fraction ($f_{\rm{Edd}} = \dfrac{L_{\rm{bol}}}{L_{\rm{Edd}}}$) as a function of \lambdasBHAR\ at $z = 0$ for each SFR classification of simulated galaxies in the UniverseMachine (as indicated by the colors in the legend) following our procedure shown in  Fig.~\ref{fig:method}. The top panel shows the derived probability distributions that result from our model (histograms), and the \citet{acg2019} probability distributions we use as constraints (smooth curves). These smooth curves are multiplied by two to account for obscured AGN (see Section~\ref{sec:method}). These two sets of data closely match one another by construction according to the procedure we use to build SMBH growth histories. The right panel shows the histograms of $f_{\rm{Edd}}$ for the different SFR classification bins we use and demonstrates that star-forming galaxies are typically assigned higher Eddington ratios than quiescent galaxies. The horizontal dotted black line denotes the Eddington limit ($f_{\rm{Edd}}=1$) and the black dashed line indicates the nominal scaling between $f_{\rm{Edd}}$ and $\lambda_{\rm{sBHAR}}$ adopted by \citet{acg2019}.
}
\label{fig:feddlambda}
\end{figure}

\subsubsection{Calculating SMBH mass growth rates}

Once the correct fraction of the galaxy population, $f_{\rm{AGN}}$, are assigned a \lambdasBHAR\ value, we then use these values to calculate \Lbol\ for every galaxy using Equation~\ref{eqn:lambda}. The calculated \Lbol\ values are directly proportional to \mstar, meaning more massive galaxies will host SMBHs with higher \Lbol\ for the same \lambdasBHAR. We then calculate the SMBH accretion rate using the equation

\begin{equation}
\dot M_{\rm{BH}} = \dfrac{(1-\eta)L_{\rm{bol}}}{\eta c^2}
\label{eqn:mdotbh}
\end{equation}
where $\eta$ is the radiative efficiency and is set to a value of $\eta = 0.1$. This value is used to calculate the \mbh\ at the previous snapshot corresponding to a timestep toward higher redshift,

\begin{equation}
M_{\rm{BH}}^{{\rm{prog}}} = M_{\rm{BH}}^{{\rm{desc}}} - \dot M_{\rm{BH}}^{{\rm{desc}}} \Delta t
\label{eqn:mbhprog}
\end{equation}
where $M_{\rm{BH}}^{{\rm{prog}}}$ is the progenitor SMBH mass being calculated, $M_{\rm{BH}}^{{\rm{desc}}}$ is the SMBH mass of the descendant, and $\Delta t$ is the size of the timestep between snapshots. Our boundary conditions are given by the \mbh\---\mstar\ relations for star-forming and quiescent galaxies at $z = 0$ (see Section~\ref{sec:mbh}). We build SMBH growth histories backwards in time, hence we subtract the amount of mass growth from the descendant \mbh\ to obtain the progenitor mass. We follow this procedure through each snapshot of the UniverseMachine, in even timesteps of $\Delta t \sim 0.1$ Gyr backward in time.

Mergers are treated by assuming the stellar mass ratios between the multiple progenitor galaxies and single descendant galaxy ($\dfrac{M_{\rm{star}}^{{\rm{prog,1}}}}{M_{\rm{star}}^{{\rm{desc}}}}$, $\dfrac{M_{\rm{star}}^{{\rm{prog,2}}}}{M_{\rm{star}}^{{\rm{desc}}}}$, ...) are the same as the SMBH mass ratios. For N number of progenitors, the SMBH mass of each progenitor is calculated using the following equation:

\begin{equation}
M_{\rm{BH}}^{{\rm{prog,N}}} = M_{\rm{BH}}^{{\rm{desc}}} \dfrac{M_{\rm{star}}^{{\rm{prog,N}}}}{M_{\rm{star}}^{{\rm{desc}}}} - \dot M_{\rm{BH}}^{{\rm{desc}}} \dfrac{M_{\rm{star}}^{{\rm{prog,N}}}}{M_{\rm{star}}^{{\rm{desc}}}} \Delta t.
\label{eqn:mdotbhmerg}
\end{equation}
We follow the growth histories of all SMBHs via accretion and mergers up until \mbh\ $< 0$ M$_{\odot}$ at which point a zero mass SMBH is assumed. For our fiducial model 23\% (26\%) of galaxies have \mbh\ $ = 0$ M$_{\odot}$ at $z = 1$ ($z = 2$). 

In Figure~\ref{fig:feddlambda} we show the Eddington fraction (${L_{\rm{bol}}}/{L_{\rm{Edd}}}$) as a function of \lambdasBHAR\ at $z = 0$. \Lbol\ is calculated based on Equation~\ref{eqn:lambda}. The systematic offset in $f_{\rm{Edd}}$ between quiescent and star-forming galaxies is due to the initial conditions that assign quiescent galaxies more massive SMBHs, and therefore systematically lower Eddington fractions, than star-forming galaxies (See Section~\ref{sec:mbh}). The histograms in the top panel show the probability distributions for \lambdasBHAR\ that we obtain from the results of our procedure. The smooth curves show the probability distributions from \citet{acg2019} multiplied by 2 to reflect the population of obscured AGN (see Section~\ref{sec:assign}). These two agree by design and serve to show how the observational data are reproduced in our model. We confirm that this agreement is reproduced at all redshifts explored in this work.

\begin{figure*}
\centering
\epsfxsize=16cm
\epsfbox{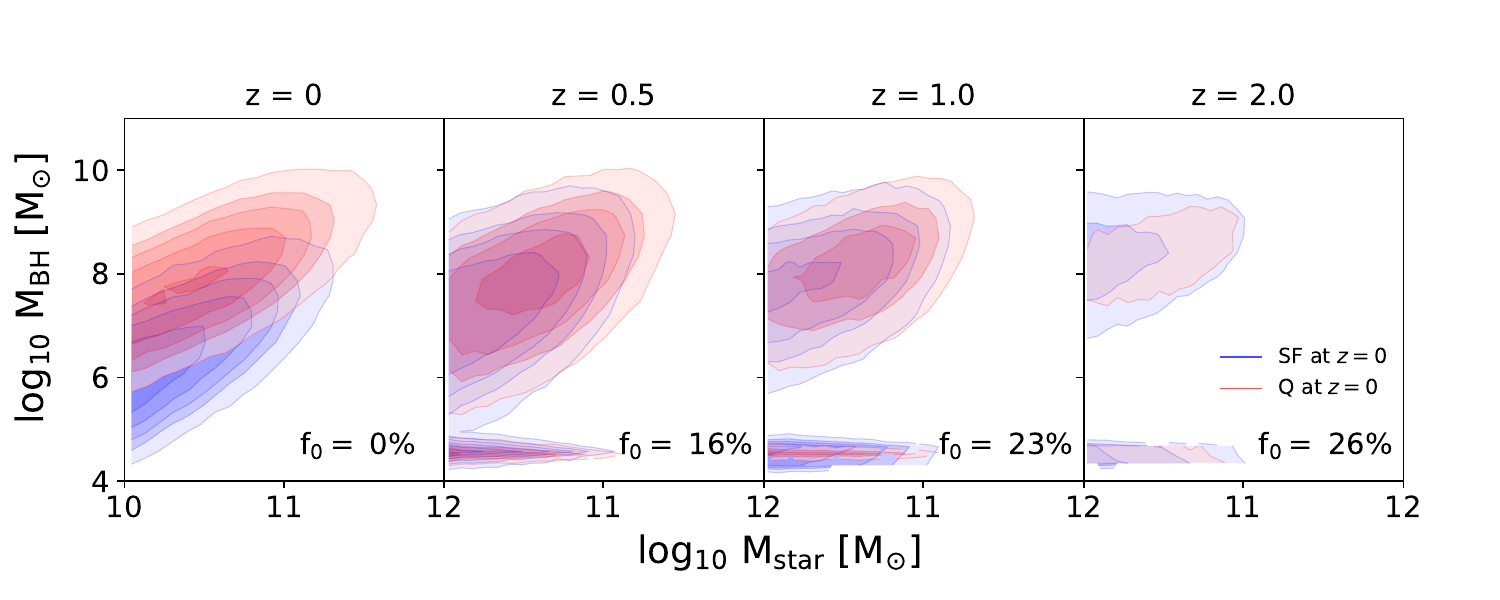}
\caption{The evolution of the \mbh\---\mstar\ relation for the fiducial model at $z = 0$, 0.5, 1, and 2. Blue and red contours show the distribution of star-forming and quiescent galaxies at the redshifts shown. There are a significant number of (primarily star-forming) galaxies whose empirically assigned SMBH accretion rates are high enough to account for all their mass growth in the late universe. These are shown as a distribution at M$_{\rm{BH}} = 10^{4.5}$ M$_{\odot}$. The fraction of \mstar\ $> 10^{10}$ M$_{\odot}$ galaxies with \mbh\ $= 0$ M$_{\odot}$, f$_0$, at a given redshift is indicated at the lower right of each panel. The distinct \mbh\---\mstar\ relation at $z = 0$ quickly disappears into an overlapping scattered distribution by $z = 0.5$ and beyond due to the statistical approach used to determinine star formation histories in the UniverseMachine.}
\label{fig:evolution}
\end{figure*}

\begin{figure*}
\centering
\epsfxsize=16cm
\epsfbox{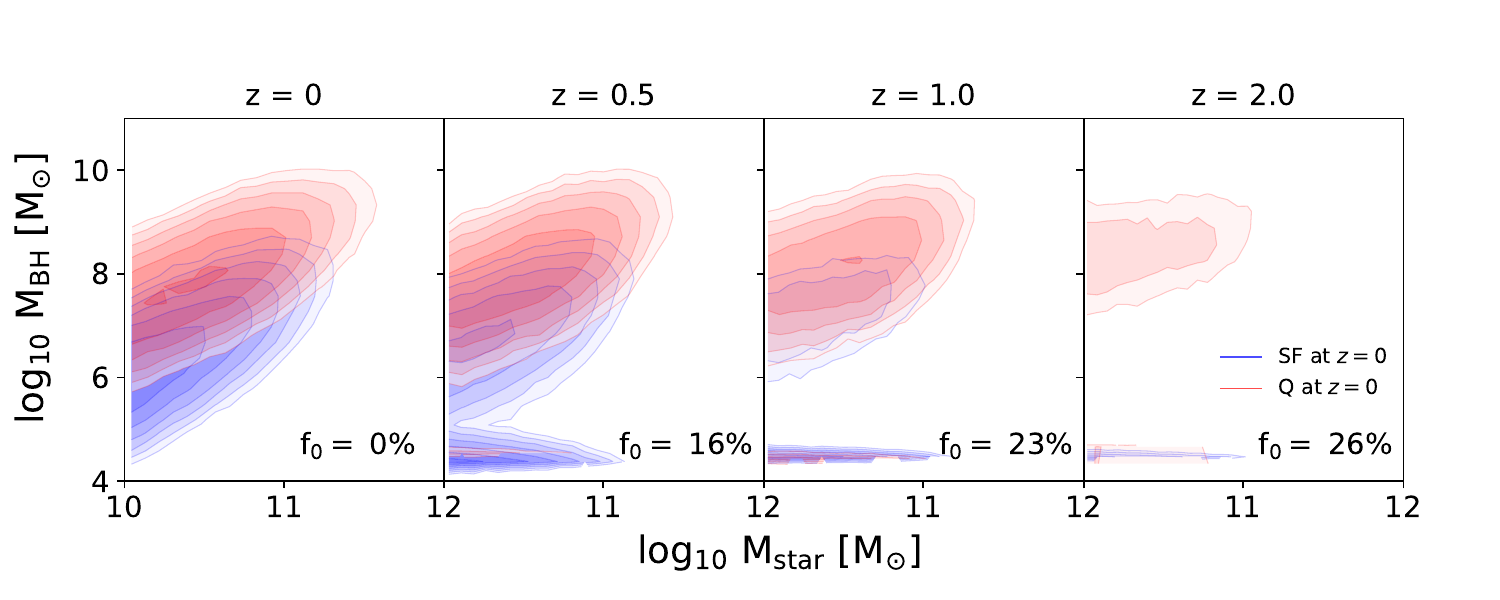}
\caption{Similar to Fig.~\ref{fig:evolution}, the evolution of the \mbh\---\mstar\ relation for the fiducial model at $z = 0$, 0.5, 1, and 2. Here, however, blue and red contours indicate the galaxy populations that end up being star-forming and quiescent at $z = 0$. Compared to Fig.~\ref{fig:evolution}, the red and blue populations remain distinct, indicating that the progenitors of $z = 0$ quiescent galaxies have hosted more massive SMBHs throughout the entirety of their histories. The progenitors of $z = 0$ star-forming galaxies have always hosted relatively undermassive SMBHs in comparison, however they largely disappear by $z = 2$. This happens either because their stellar masses become $< 10^{10}$ M$_{\odot}$ at higher redshift or their SMBH masses have grown all of their mass at late times, leaving them with \mbh\ $= 0$ M$_{\odot}$. The fraction of galaxies with \mbh\ $= 0$ M$_{\odot}$, f$_0$, is the same as the values shown in Fig.~\ref{fig:evolution}.}
\label{fig:forecast}
\end{figure*}


\section{Evolution of the \mbh\---\mstar\ relation}
\label{sec:scalingrelations}

Fig.~\ref{fig:evolution} shows the \mbh\---\mstar\ relations for the full population of galaxies in the UniverseMachine at $z = 0$, 0.5, 1, and 2. We plot star-forming and quiescent galaxies in blue and red at each redshift shown. The leftmost plot shows the relations introduced by our modeling approach at $z = 0$ while the subsequent plots show the results of our SMBH evolutionary growth model.


The fraction of galaxies with \mbh\ $= 0$ \msun\ is indicated in the bottom right of each panel ($f_0$). At $z = 0$, all galaxies are assigned a SMBH mass. The presence of a population of galaxies with no SMBHs at higher redshifts indicates that our model's assignment of $\dot{\rm{M}}_{\rm{BH}}$ for these galaxies has accounted for all of their SMBH mass growth by the redshift shown. We show the galaxies with no SMBHs on the main panels as a distribution of upper limits around the value log$_{10}$ \mbh\ $\sim 4.5$. The galaxies with no SMBHs continue to exist in the simulation at higher redshifts and may evolve to have lower stellar masses $< 10^{10}$ \msun\ that are outside of the axes limit of our figures and that are not tracked by our analysis in this paper.

We find that the \mbh\---\mstar\ relation evolves for both the quiescent and star forming populations. These two populations have different \mbh\---\mstar\ relations at $z = 0$ by construction, but by $z = 0.5$ these two populations converge to have very similar, overlapping distributions. The similarity in SMBH mass distributions persists out to $z = 2$.

Several studies find that star formation histories in the UniverseMachine are substantially more variable than in other models \citep{itg2020, ahb2023}. This is a result of their statistical approach; the UniverseMachine does not have any \textit{physical} mechanism for quenching included in their model. It is a statistical model that determines a galaxy's SFR from its halo assembly history. We find that the majority of galaxies in the UniverseMachine show bursty star formation histories that result in substantial mixing between star-forming and quiescent populations on the \mbh\---\mstar\ relation at $z > 0$ (see Fig.~\ref{fig:evolution}). While the ensemble population's growth histories produce a growing quiescent population with decreasing redshift \citep{bwh2019}, we find that classifying galaxies by their SFR at each redshift produces more mixing between populations than would be the case for a more physically-motivated model, due to the stochastic and statistical nature of the UniverseMachine. We discuss this further in Section~\ref{sec:caveats}.



For this reason we decide for the remainder of this paper to show galaxies as star-forming or quiescent based on their $z = 0$ classification. Figure~\ref{fig:forecast} shows the same data as Figure~\ref{fig:evolution} except instead of showing the populations split by their star-forming/quiescent classification at the redshift shown, the split in color indicates the galaxies that are star-forming/quiescent at $z = 0$.

These panels show substantially less mixing between galaxy populations with increasing redshift. Galaxies that will become quiescent by $z = 0$ always have higher SMBH masses at all redshifts out to $z = 2$. Galaxies that are star-forming at $z = 0$ make up the undermassive envelope and their evolution results in the gradual removal of these galaxies from the relation with increasing redshift. Much of this latter population reaches stellar masses $< 10^{10}$ \msun\ (below the stellar masses we explore in this work) but a substantial fraction also make up the population of galaxies with \mbh\ $= 0$ \msun. The fraction of all galaxies that have \mbh\ $= 0$ \msun, f$_0$, are the same as those indicated in Fig.~\ref{fig:evolution}.

\begin{figure*}
\centering
\epsfxsize=7cm
\epsfbox{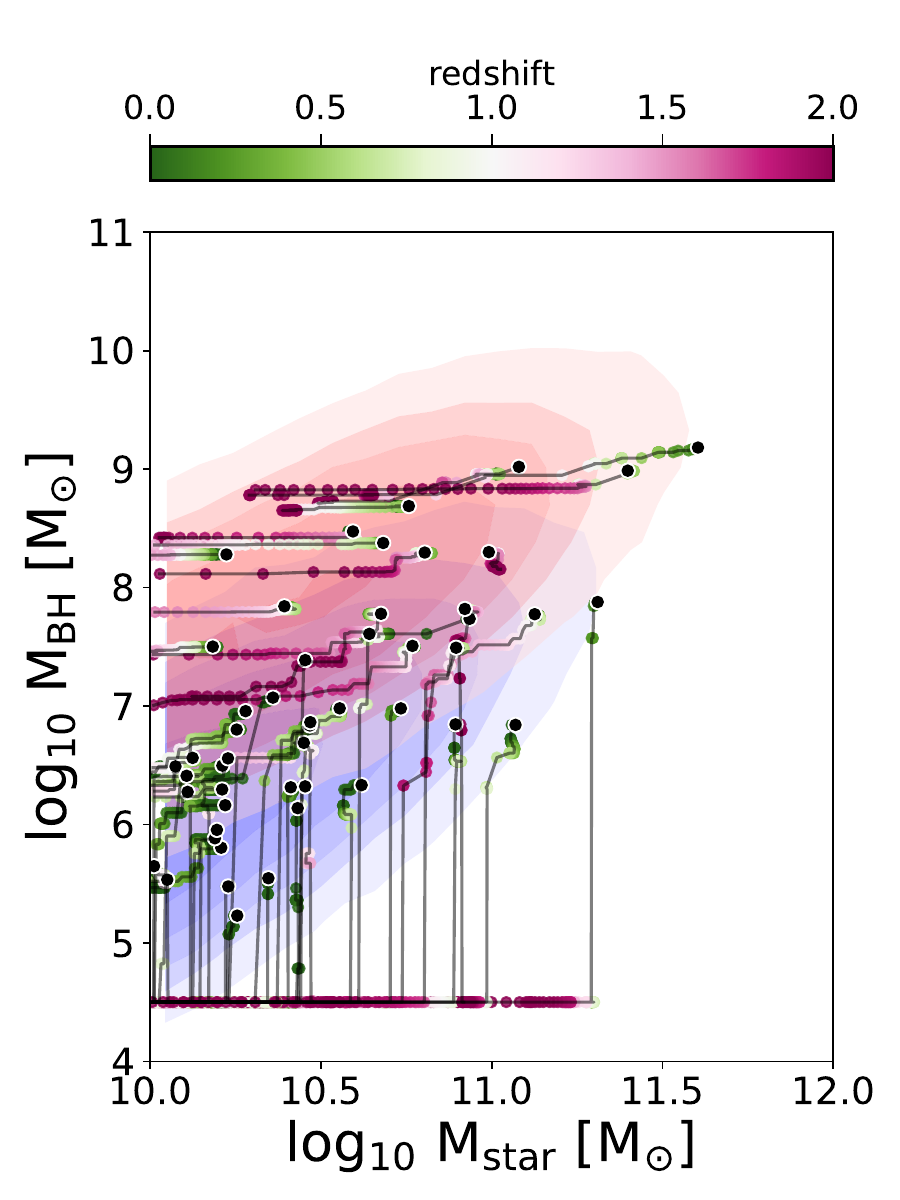}
\epsfxsize=7.6cm
\epsfbox{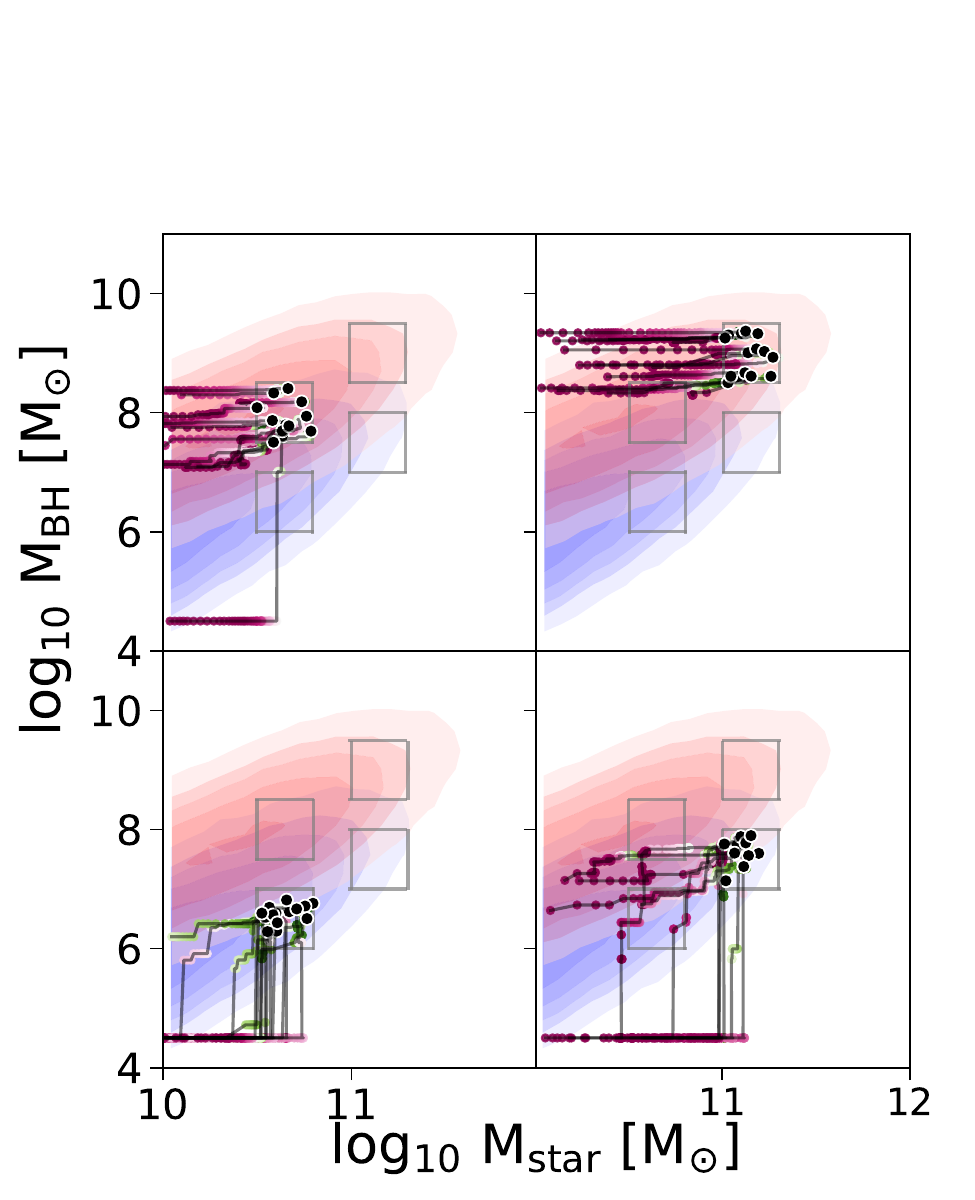}
\caption{{\it Left panel:} Individual galaxy growth histories on the \mbh\---\mstar\ relation. The $z = 0$ distributions of star-forming and quiescent galaxies are shown in blue and red contours. Black data points indicate a subset of the UniverseMachine galaxy population at $z = 0$. The growth histories of this subset are indicated by black lines with colored points indicating their position on this plot with increasing redshift on a green to pink color map. Quiescent galaxies with overmassive SMBHs at $z = 0$ have a more active stellar mass growth history at higher redshift (pink) whereas star-forming galaxies with undermassive SMBHs at $z = 0$ show more growth in both stellar and SMBH mass at lower redshift (green). {\it Right panels:} Growth histories for $z = 0$ galaxies that host overmassive (top) and undermassive (bottom) SMBHs with $z = 0$ stellar masses in two different ranges (left and right panels). Galaxies with overmassive SMBHs at $z = 0$ have more horizontal tracks and grow their SMBH and stellar mass early whereas star-forming galaxies at $z = 0$ have steeper tracks and grow their SMBH and stellar mass at late times.}
\label{fig:tracks}
\end{figure*}

Regardless of how we show star-forming or quiescent galaxies on these plots the scatter in \mbh\ at a given \mstar\ is substantial at all redshifts for both populations. The relations for these two populations show a decrease in slope and increase in normalization from redshift $z = 0$ to $z = 2$. This results from the diversity of growth histories for galaxies that lie at one locus on the \mbh\---\mstar\ plot at $z = 0$. Galaxies with overmassive SMBHs that end up with similar \mbh\ and \mstar\ values diverge when some of those galaxies grow their stellar mass more substantially while others grow more slowly. All the while, lower mass SMBHs have a higher likelihood of falling off the relation and joining the population with \mbh\ $= 0$ \msun. Comparing the $z = 2$ panels in Fig.~\ref{fig:evolution} and~\ref{fig:forecast}, we demonstrate that in our model the $z = 2$ \mbh\---\mstar\ relation above \mstar\ $> 10^{10}$ \msun\ is almost entirely made up of galaxies that will become quiescent by $z = 0$ (red contours in Fig.~\ref{fig:forecast}), despite being mostly star-forming at $z = 2$ (blue contours in Fig.~\ref{fig:evolution}).



Fig.~\ref{fig:tracks} shows individual galaxy tracks in the \mbh\---\mstar\ plane, showing the diversity of growth histories for SMBHs alongside their stellar mass growth. The right panels show the growth histories of galaxies occupying different loci on the \mbh\---\mstar\ relation at $z = 0$. The black circles are the initial (\mbh, \mstar) values that these galaxies have at $z = 0$ before tracking their growth histories backwards in time to lower masses.

Galaxies that end up quiescent and have higher \mbh/\mstar\ ratios at $z = 0$ exhibit very little fractional growth in SMBH mass and generally have flat, horizontal tracks in this space, indicating substantially more stellar mass growth than SMBH growth since $z = 2$. This stellar growth primarily occurs at early times, indicated by the largely pink evolutionary tracks. Galaxies that end up as star-forming with lower \mbh/\mstar\ ratios at $z = 0$ have more varied growth histories, growing in both stellar and SMBH masses. A substantial fraction of galaxies with low \mbh\ and low \mstar\ have more vertical tracks, showing rapid SMBH growth at late times that can account for all of the $z = 0$ accumulated SMBH mass. These galaxies often show periods of rapid SMBH mass growth interchanged between periods of little or no growth, indicating the stochasticity of accretion events in our model. This feature comes about due to assigning only a fraction of galaxies in each SFR bin, $f_{\rm{AGN}}$, an accretion rate at any one snapshot in the model.


 
\begin{figure}
\centering
\epsfxsize=7.5cm
\epsfbox{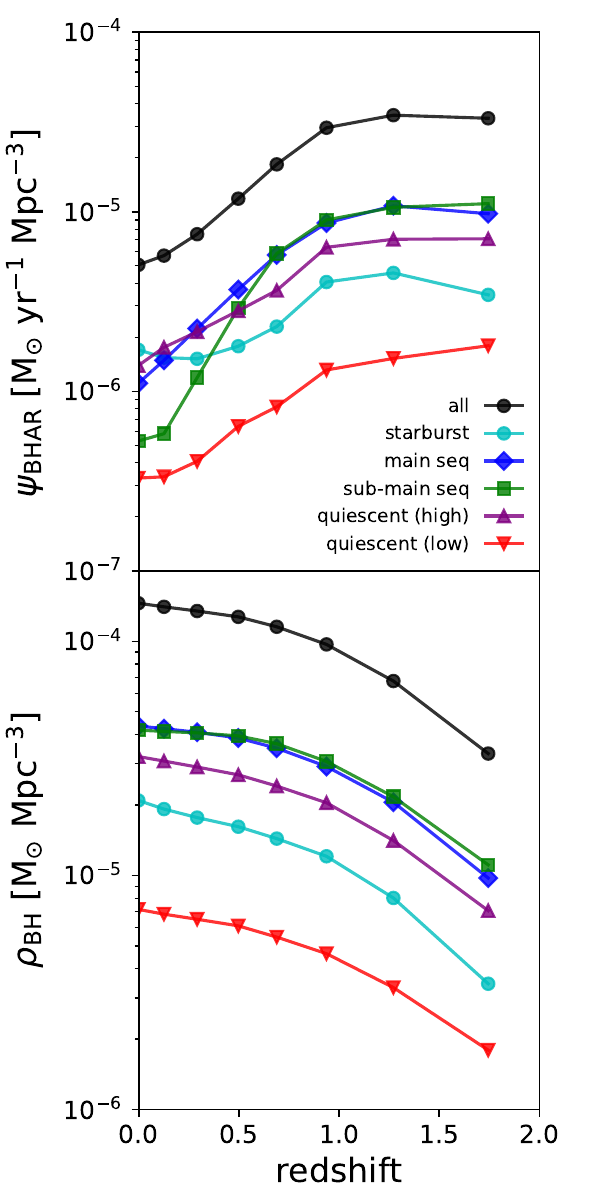}
\caption{\textit{Top panel:} The total SMBH mass growth rate per unit Mpc$^{3}$ as a function of redshift for each SFR classification phase (colors) and for all bins together (black). \textit{Bottom panel:} The cumulative version of the top panel, showing the growth of SMBH mass over time.}
\label{fig:growthdensity}
\end{figure}

\section{In what phase of the galaxy lifecycle do SMBHs grow?}
\label{sec:phase}

The top panel of Figure~\ref{fig:growthdensity} shows the SMBH mass growth rate per unit volume as a function of redshift for each of our SFR classification bins (colored data) and the total growth from all bins (black data). These results were compiled by integrating the total SMBH mass growth during each phase for all galaxies with \mstar\ $> 10^{10}$ M$_{\odot}$ in a given population and dividing by the time between the datapoints shown. The bottom panel shows the cumulative version of the top panel, indicating the total amount of SMBH mass per cubic Mpc in our model as a function of redshift.

While the UniverseMachine has too much mixing between populations with different SFRs due to their bursty star formation histories for individual galaxies (see Section~\ref{sec:scalingrelations} and~\ref{sec:caveats}), the SFR-\mstar\ relations for the \textit{overall} populations at each timestep in their model are by construction made to agree with observational constraints (see Section~\ref{sec:um}). For this reason, our analysis in Fig.~\ref{fig:growthdensity} is robust despite the bursty star formation histories, as it depends on the UniverseMachine model producing the accurate number of galaxies and their stellar mass distribution in each of our SFR classification bins.


Between $z = 0.5-2$, most SMBH mass growth occurs in the main sequence and sub-main sequence phases, followed by the quiescent (high), starburst, and quiescent (low) phases, respectively. At lower redshift between $z = 0-0.5$, the \mbh\ growth rate in the quiescent (high) phase increases to slightly exceed the growth rate in the main sequence and sub-main sequence phases. The growth rate in the sub-main sequence phase decreases to about half the growth in the main sequence phase at late cosmic times. The flattening of the starburst curve in the top panel at $z < 0.5$ is likely due to the low numbers of starburst galaxies found in the observational data at low redshifts. The \lambdasBHAR\ distributions for the low redshift starburst population are therefore less reliable and the sBHARs we assign are more uncertain at these low redshifts for this particular population.

\begin{figure*}
\centering
\epsfxsize=5.6cm
\epsfbox{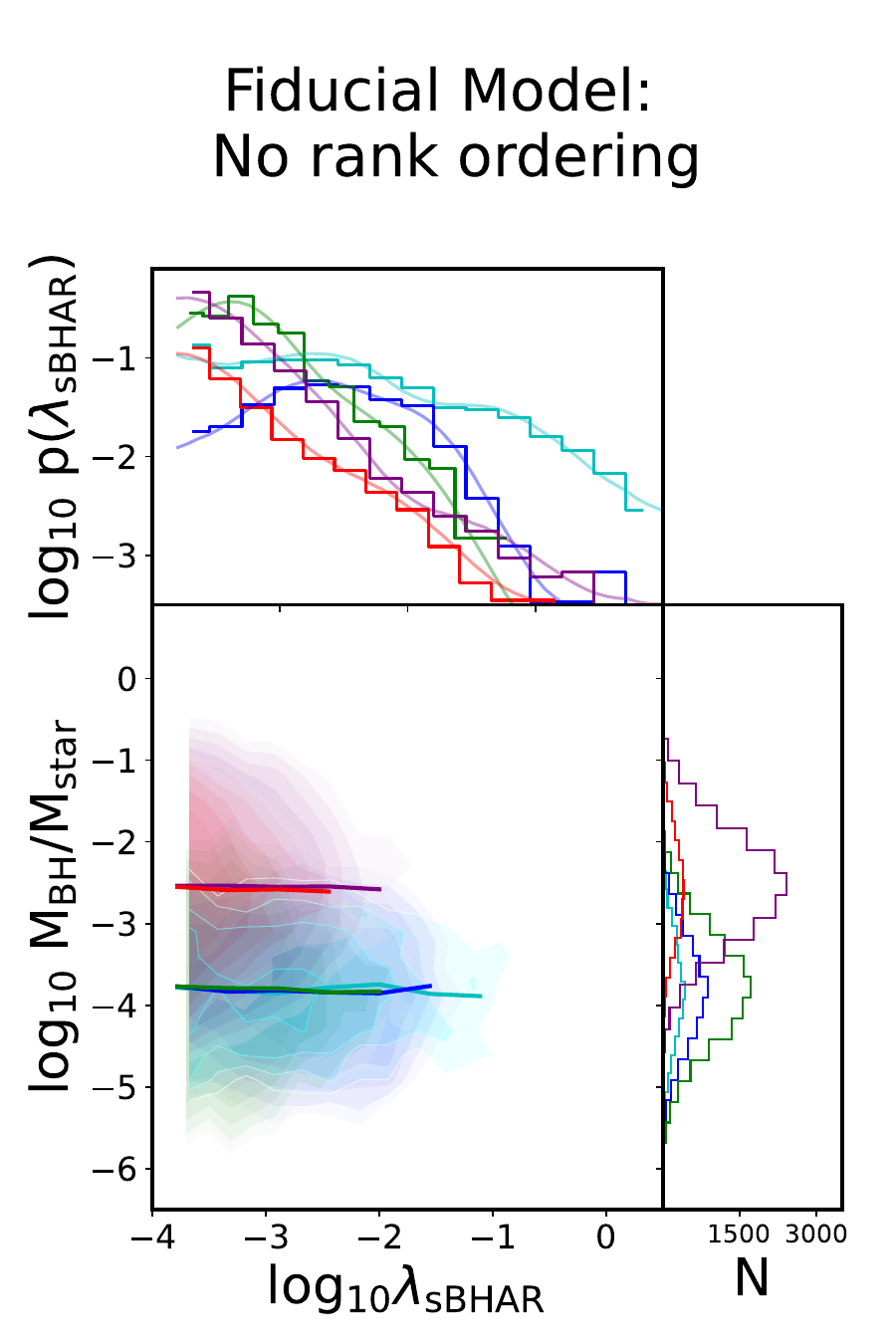}
\epsfxsize=5.6cm
\epsfbox{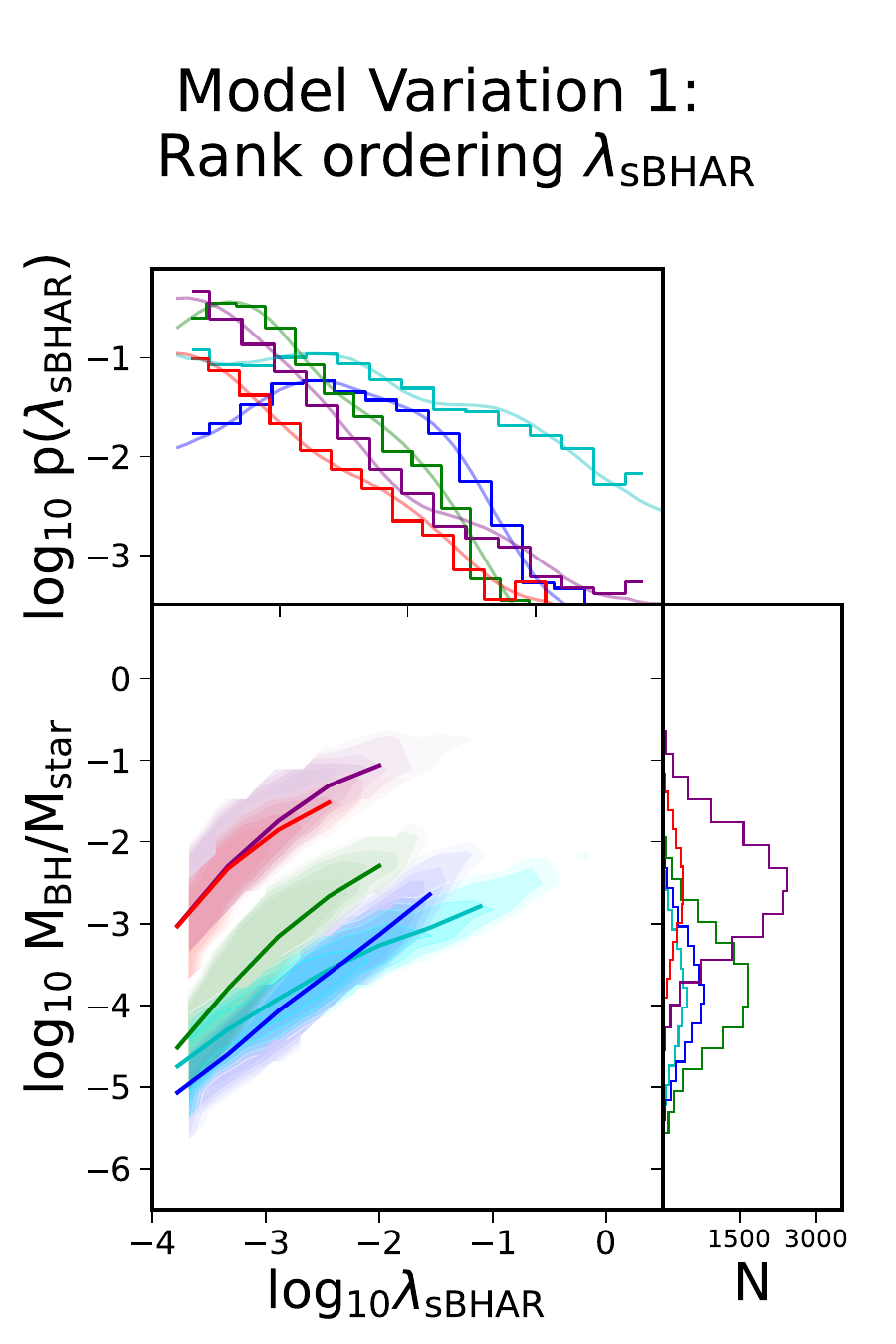}
\epsfxsize=5.6cm
\epsfbox{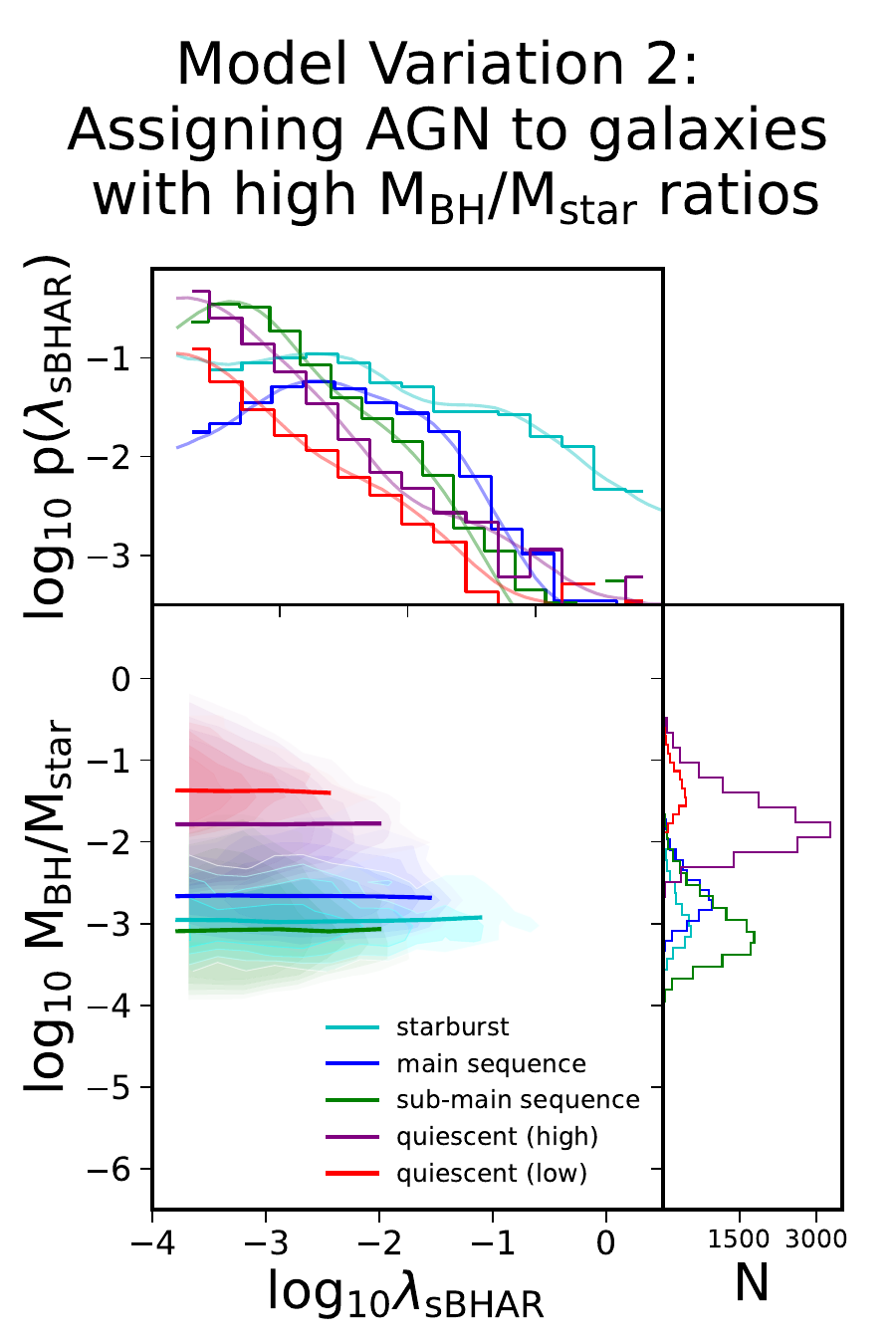}
\caption{The \mbh\---\mstar\ ratio as a function of \lambdasBHAR\ at $z = 0$ for the fiducial model and Model Variations 1 and 2, similar to Figure~\ref{fig:feddlambda}. The top panels show the derived probability distributions that result from our model (histograms), and the \citet{acg2019} probability distributions (smooth curves). These two curves for all models closely match one another by construction, regardless of the differences in how we assign the \lambdasBHAR\ values in the model variations. The right panel shows the histograms of \mbh\---\mstar\ ratio for the different SFR classification bins we use. Both Model Variations 1 and 2 are designed to preferentially assign high accretion rates to galaxies with overmassive SMBHs as test cases for our model.}
\label{fig:mvs}
\end{figure*}

\begin{figure*}
\centering
\epsfxsize=16cm
\epsfbox{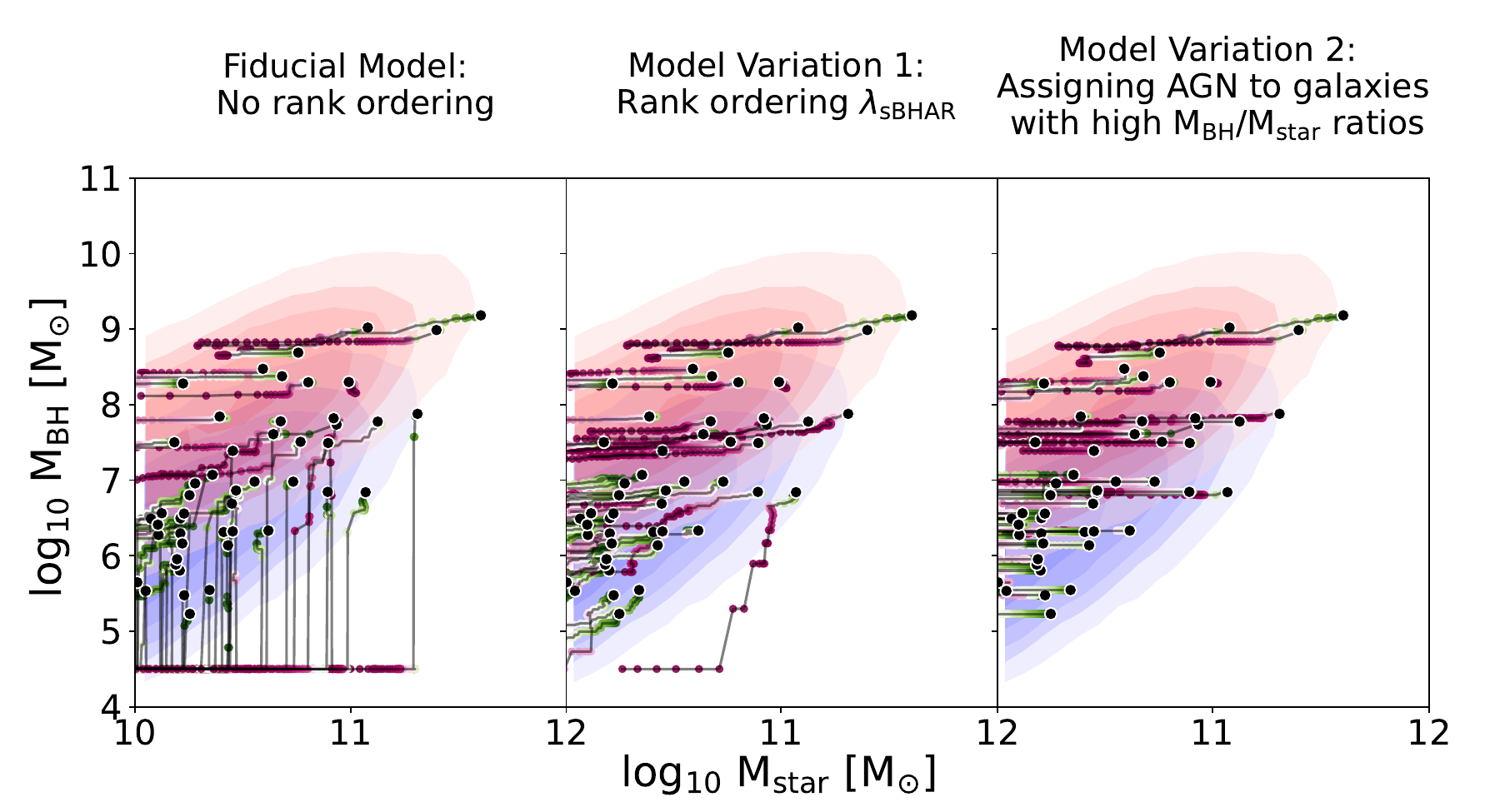}
\caption{Individual galaxy growth histories on the \mbh\---\mstar\ relation for our fiducial model (left panel), Model Variation 1 (center panel) and Model Variation 2 (right panel), similar to Figure~\ref{fig:tracks}. The $z = 0$ distribution of star-forming and quiescent galaxies is shown in blue and red contours in the background. The growth histories are indicated by lines that stem from black  circles and are colored by increasing redshift on a green to pink color map. Model Variations 1 and 2 preferentially assign high accretion rates to galaxies with overmassive SMBHs. Despite this intentional bias, we find that these massive SMBHs grow very little compared to their already very high SMBH masses.}
\label{fig:tracksmvs}
\end{figure*}

At all redshifts it is clear that quiescent galaxies with lower SFRs (red data) do not substantially contribute to the growth of SMBHs. However, we find that those in the quiescent (high) bin grow a substantial amount of SMBH mass. Surprisingly, this exceeds the amount grown during the starburst phases. Galaxies that undergo a starburst phase generally grow more \mbh\ during one timestep than those in the quiescent (high) phase during the same amount of time. However, galaxies pass through the starburst phase less frequently than the quiescent (high) phase, resulting in less total \mbh\ growth. \citet{ack2022} find that a substantial amount of \mbh\ growth occurs in extended quiescent galaxies due to the high AGN fractions in these systems. The substantial growth in the quiescent (high) phase in our model reflects the importance of \mbh\ growth in quiescent galaxies despite a lack of active star-formation.

Although most star-forming galaxies lie on the main sequence, a similar amount of SMBH mass growth occurs in the less populated sub-main sequence phase at $z > 0.5$ \citep{acg2019} but quickly drops at lower redshifts. Other observational studies confirm the lack of AGN in sub-main sequence galaxies at low redshifts \citep{bwa2022, bwa2023}.

We find that the total \mbh\ growth rate density decreases with decreasing redshift, in agreement with observational studies of BHAR density evolution based on X-ray data \citep{mh2013, dgp2014}. Our work is novel in determining how much of this BHAR occurs during different phases of star formation activity within the galaxy population. Calculations from observations generally do not make a \mstar\ cut, therefore we may slightly underestimate the total BHAR in the universe in our model.


\section{Inefficient growth of the most massive SMBHs}
\label{sec:highMbh}

We found in Section~\ref{sec:scalingrelations} that growing higher mass SMBHs above $\sim10^{8}$ M$_{\odot}$ is difficult and their evolutionary tracks on the \mbh\---\mstar\ relation are mostly horizontal. In this section we run two different versions of our simulations to determine whether assigning more accretion growth to the most massive SMBHs would change this behavior. In Model Variation 1 we rank order galaxies based on their  \mbh/\mstar\ ratios, preferentially assigning high \lambdasBHAR\ to galaxies with overmassive SMBHs for their host galaxy stellar mass. In Model Variation 2 we only assign \lambdasBHAR\ values to galaxies with high \mbh/\mstar\ ratios. Fig.~\ref{fig:mvs} shows the \mbh/\mstar\ ratio as a function of \lambdasBHAR\ for the fiducial model and both Model Variations 1 and 2 to demonstrate the differences between these models. The top panels show the probability distributions produced by our model (histograms) and those from \citet{acg2019} (same as the top panel of Fig.~\ref{fig:feddlambda}). In all model variations these probability distributions agree by construction (see Section~\ref{sec:method}) and yet we are able to use different assumptions when assigning \lambdasBHAR\ values to galaxies. We take advantage of these degeneracies to explore various plausible assumptions about how SMBHs grow within their host galaxies.

In Model Variation 1 a \lambdasBHAR\ value is assigned to a fraction, $f_{\rm{AGN}}$, of galaxies randomly in each SFR bin, just as in the fiducial model. All galaxies have an equal probability of hosting an AGN (i.e., being assigned a non-zero \lambdasBHAR\ value) but the difference is that galaxies with higher \mbh/\mstar\ ratios are now rank ordered and preferentially assigned a higher \lambdasBHAR\ value. In Model Variation 2 only galaxies with high \mbh\---\mstar\ ratios are assigned a \lambdasBHAR, meaning that only these galaxies have the possibility of hosting an AGN in this version. This last model is extreme and likely does not correspond to how SMBHs grow in the real Universe. However our goal in implementing this version is to create a strong limiting test case. 

\begin{figure*}
\centering
\epsfxsize=5cm
\epsfbox{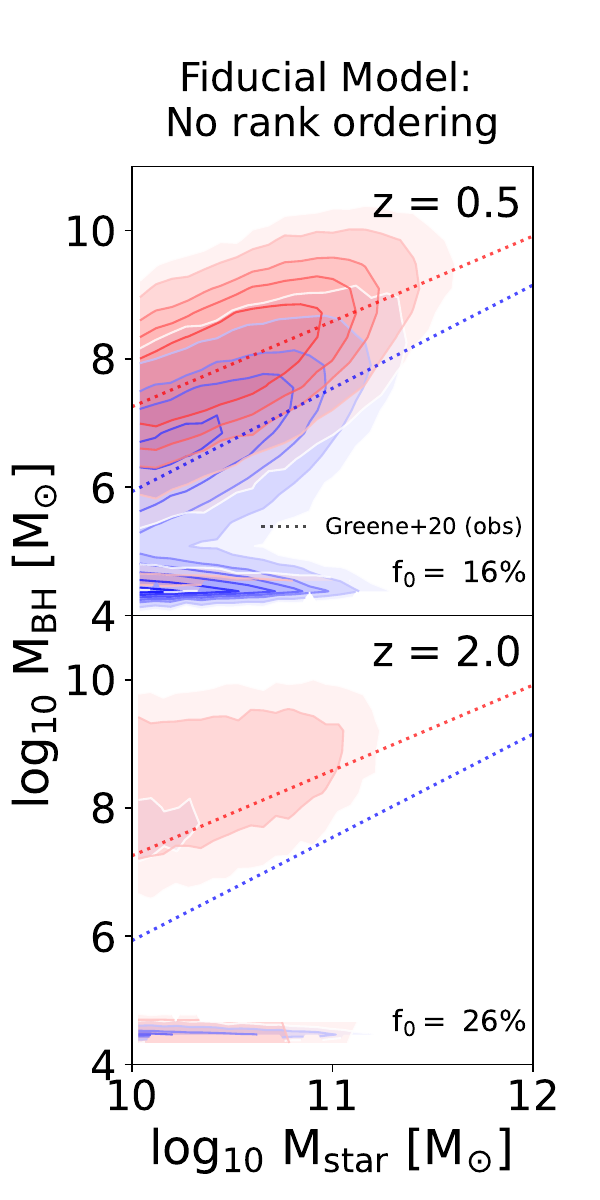}
\epsfxsize=5cm
\epsfbox{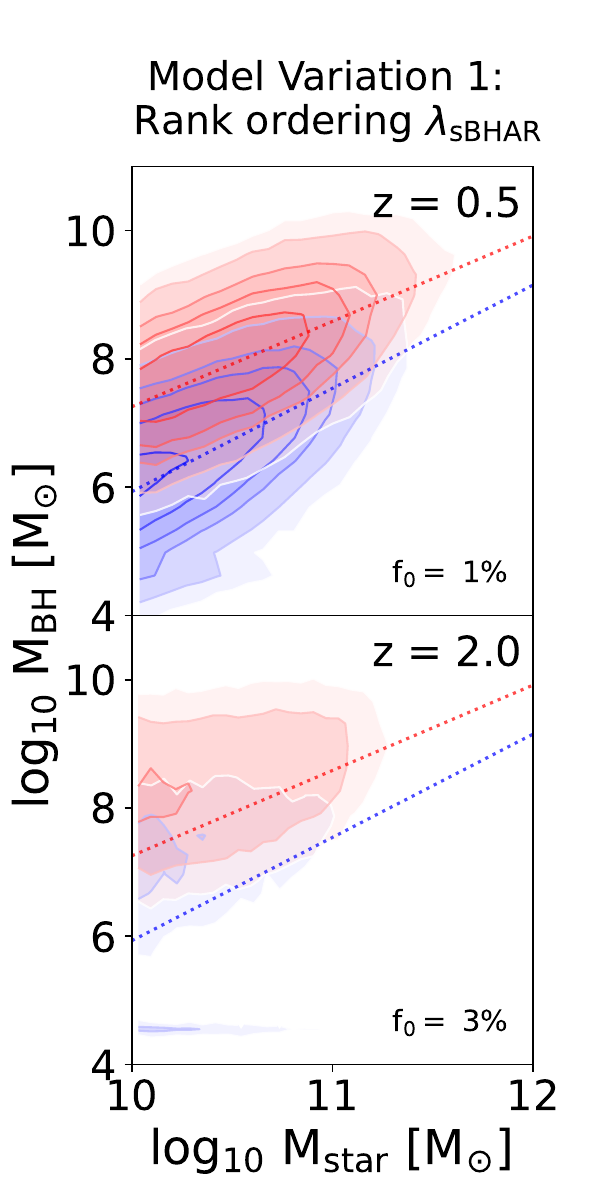}
\epsfxsize=5cm
\epsfbox{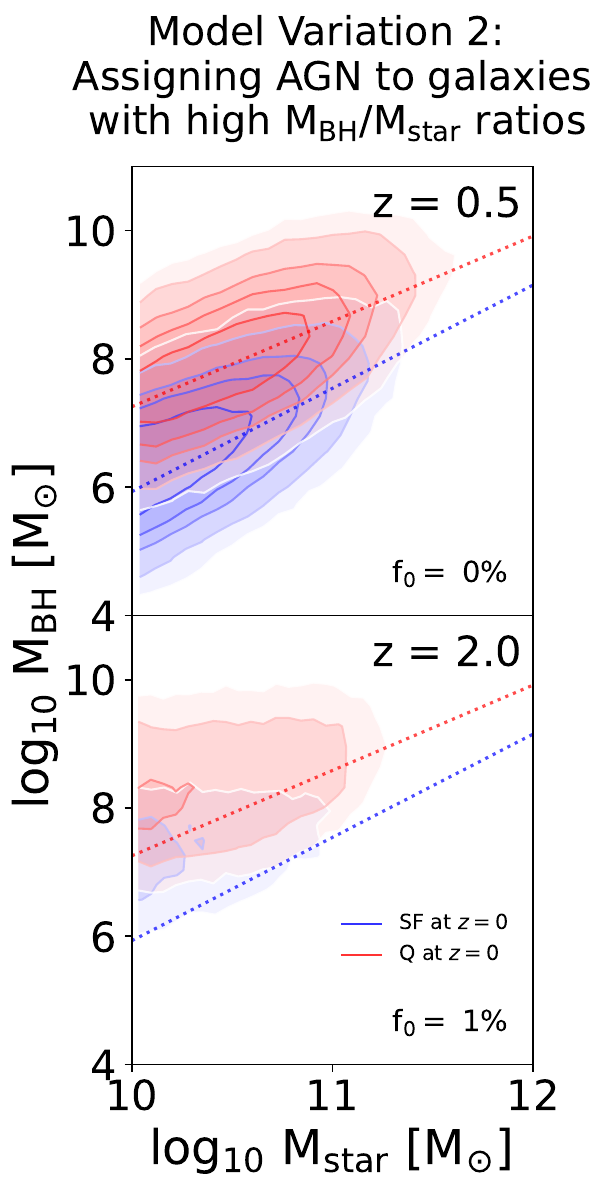}
\caption{The $z = 0.5$ and $z = 2$ \mbh\---\mstar\ relations for the fiducial model and Model Variations 1 and 2, similar to Figure~\ref{fig:evolvemvs}. The  $z = 0$ \citep{gsh2020} relations are shown by the dotted lines for comparison in every panel. The differences in the method for assigning SMBH accretion rates produce similar relations at higher redshifts for all models. The primary difference is that there are slightly more $z = 0$ star-forming galaxies that make up the relation at $z = 2$. In the fiducial model, these galaxies grew their SMBHs late and at higher redshifts made up the fraction of galaxies with \mbh\ $= 0$ M$_{\odot}$, $f_0$.}
\label{fig:evolvemvs}
\end{figure*}

Both Model Variations 1 and 2 are designed to assign higher \lambdasBHAR\ values to galaxies with overmassive SMBHs. These adjustments to the assumption of our model are made to determine whether these decisions could account for the mass growth at redshifts $< 2$ of higher mass SMBHs that are observed in the local Universe. Fig.~\ref{fig:tracksmvs} shows individual growth histories for the same subset of galaxies in each model variation across the \mbh\---\mstar\ relation. In every panel the most massive SMBHs still grow very little and show horizontal growth histories on this relation, indicating significantly more stellar mass growth than SMBH mass growth between $z = 0-2$. We conclude that preferentially assigning most of the observed accretion growth between $z = 0-2$ to the most overmassive SMBHs is not enough to account for their masses. Recent work by \citet{gag2024}, measuring the extent of SMBH growth in massive galaxy populations, reached a similar conclusion and is discussed further in Section~\ref{sec:discussion}.

The only substantial difference in growth histories between these models occurs for the star-forming population where SMBHs grow much less than in the fiducial model. In Model Variation 1 they are preferentially assigned lower \lambdasBHAR\ values, which results in lower \mbh\ growth rates as indicated by the slightly flatter growth histories in the middle panel. In Model Variation 2 they are very unlikely to be assigned any \lambdasBHAR\ at all, indicated by the completely flat tracks at low \mbh\---\mstar\ ratios (lower right portion of the \mbh\---\mstar\ relation) in the right panel. While these latter two models do not reflect observational evidence that AGN are more likely found in star-forming galaxies, they indicate the difficulty in growing the most massive SMBHs at $z < 2$.

Figure~\ref{fig:evolvemvs} shows the \mbh\---\mstar\ relations for all three models at $z = 0.5$ (top) and $z = 2.0$ (bottom) for galaxies that are star-forming (blue) and quiescent (red) at $z = 0$. The observed relations for star-forming and quiescent galaxies from \citet{gsh2020} that we use to assign SMBH masses at $z = 0$ are shown as red and blue dashed lines for comparison in all panels (see Section~\ref{sec:mbh}). We also include galaxies that have \mbh\ $= 0$ M$_{\odot}$ on the figures as a distributions around the value log$_{10}$ \mbh\ $\sim 4.5$. There is surprisingly little difference shown in this plot between these three models. The relations for the $z = 0$ quiescent population converge to be almost the same at $z > 0.5$ in all three models. This feature of the fiducial model was discussed in Section~\ref{sec:scalingrelations}. The $z = 0$ star-forming population makes up a very small fraction of the $z = 2$ relation for our fiducial model. Model Variations 1 and 2 show a slightly higher number of progenitors of the $z = 0$ star-forming population at $z = 2$ (blue contours) since their SMBH growth in the fiducial model was reallocated to galaxies that have overmassive SMBHs. These model variations therefore have low numbers of galaxies with \mbh\ $= 0$ M$_{\odot}$, indicated by lower $f_0$ percentages.

These results demonstrate that galaxies with low \mbh/\mstar\ ratios at $z = 0$ (observed as the star-forming population at a lower relative normalization) rarely appear on the $z = 2$ relation according to our model and the variations we present. This is because these galaxies have either (1) grown most of their SMBH masses at later times and have \mbh\ $= 0$ M$_{\odot}$ at higher redshifts (fiducial model), or (2) assembled a significant amount of stellar mass between $z = 0-2$ that has evolved them to have stellar masses $< 10^{10}$ M$_{\odot}$ that lie outside of the stellar mass range we study in this work (Model Variations 1 and 2). 

Since galaxies with high \mbh\ show flat growth histories despite our attempts at biasing SMBH mass growth to encourage their assembly between $z = 0-2$, these galaxies do not end up filling in the area of the \mbh\---\mstar\ relation where star-forming galaxies reside at low redshifts. Instead, their lack of SMBH growth at $z < 2$ causes the normalization of \mbh\---\mstar\ relation to increase and the slope to decrease with increasing redshift. The slope of the relation decreasing likely indicates the variety of stellar mass growth histories at earlier times for galaxies with high mass SMBHs. Those that grow very little remain at the top right of the plot, whereas those that grow more substantially at earlier times populate the upper left, flattening the $z = 2$ relation.

\begin{figure*}
\centering
\epsfxsize=15cm
\epsfbox{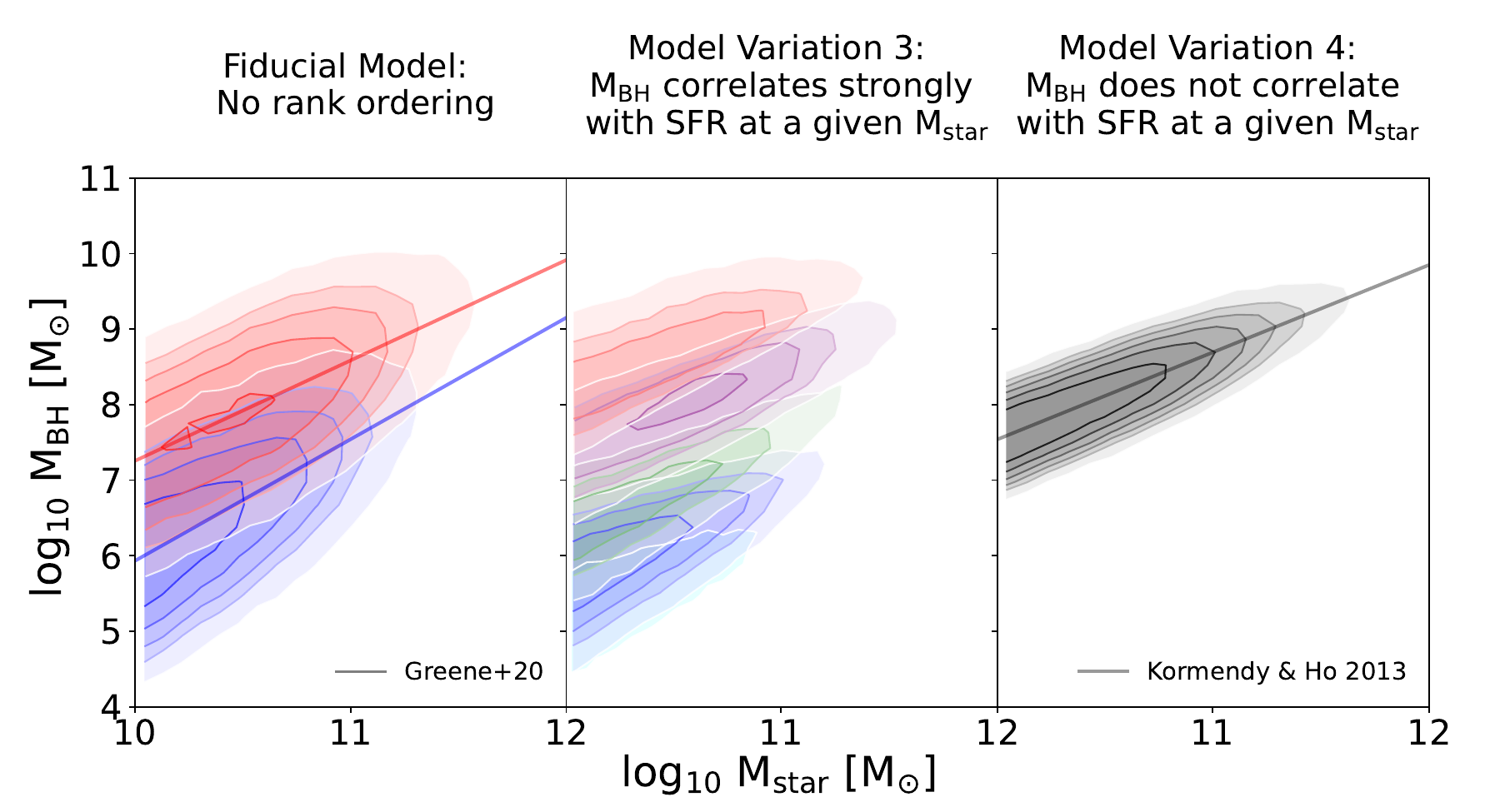}
\caption{The \mbh\---\mstar\ relations at $z = 0$ for the fiducial model (left panel), Model Variation 3 (center panel), and Model Variation 4 (right panel). Model Variation 3 assumes a strong correlation between \mbh, \mstar, and SFR at $z = 0$. Model Variation 4 assumes a tight correlation between \mbh\ and \mstar\ with no correlation with SFR at $z = 0$. This last model is a typical point of comparison for simulation results, despite its limitations in accounting for the population of undermassive SMBHs in star-forming galaxies.}
\label{fig:mbhmstarmvs}
\end{figure*}


\section{Varying the relationship between \mstar, \mbh, and SFR}
\label{sec:varrel}



A common practice for calibrating the SMBH growth model in galaxy formation simulations is to use a single \mbh\ scaling relation for all galaxies, typically the \mbh\---M$_{\rm{bulge}}$, \mbh\---\mstar\, or sometimes the \mbh\---$\sigma$ relation \citep{shc2008, hwt2015, scb2015, sdh2015, vdp2016}. The general aim is to produce simulated SMBHs that resemble observed samples at $z = 0$. However, much progress has been made in understanding the correlations between SMBHs and their host galaxies beyond a simple single scaling relation. For instance, many studies indicate that star-forming and quiescent galaxies lie on different regions of the \mbh\---\mstar\ relation at $z = 0$ \citep{rv2015, tbh2016, gsh2020}. These studies form the basis for our $z = 0$ assignment of SMBH masses in our empirical model for SMBH growth (see Section~\ref{sec:mbh}). 

In our fiducial model we assign SMBH masses based on whether a galaxy is star-forming or quiescent, using the different relations observed in the local universe for these two populations \citep{gsh2020}. In this section we alter these assumptions and run two additional model variations that differ in the assignment of \mbh\ at $z = 0$ (see Section~\ref{sec:mbh}). In Model Variation 3 we assign galaxies a \mbh\ value assuming the ratio of \mbh\ to \mstar\ is a strong function of SFR. This produces a maximally SFR-dependent version of our fiducial model and uses the relation between \mbh, \mstar, and SFR identified by \citet{tbw2017} for the observed $z = 0$ population of galaxies with dynamically-measured SMBH masses (see Eqn.~\ref{eqn:t17}). In Model Variation 4 we assign galaxies a \mbh\ value assuming all galaxies lie on the same \mbh\---\mstar\ relation given by \citet{kh2013}, regardless of their host galaxy SFR (see Eqn.~\ref{eqn:kh13}). This model assumes the opposite extreme as Model Variation 3 and mimics the traditional assumptions used to calibrate and test state-of-the-art galaxy formation models.

\begin{figure*}[h]
\centering
\epsfxsize=5.6cm
\epsfbox{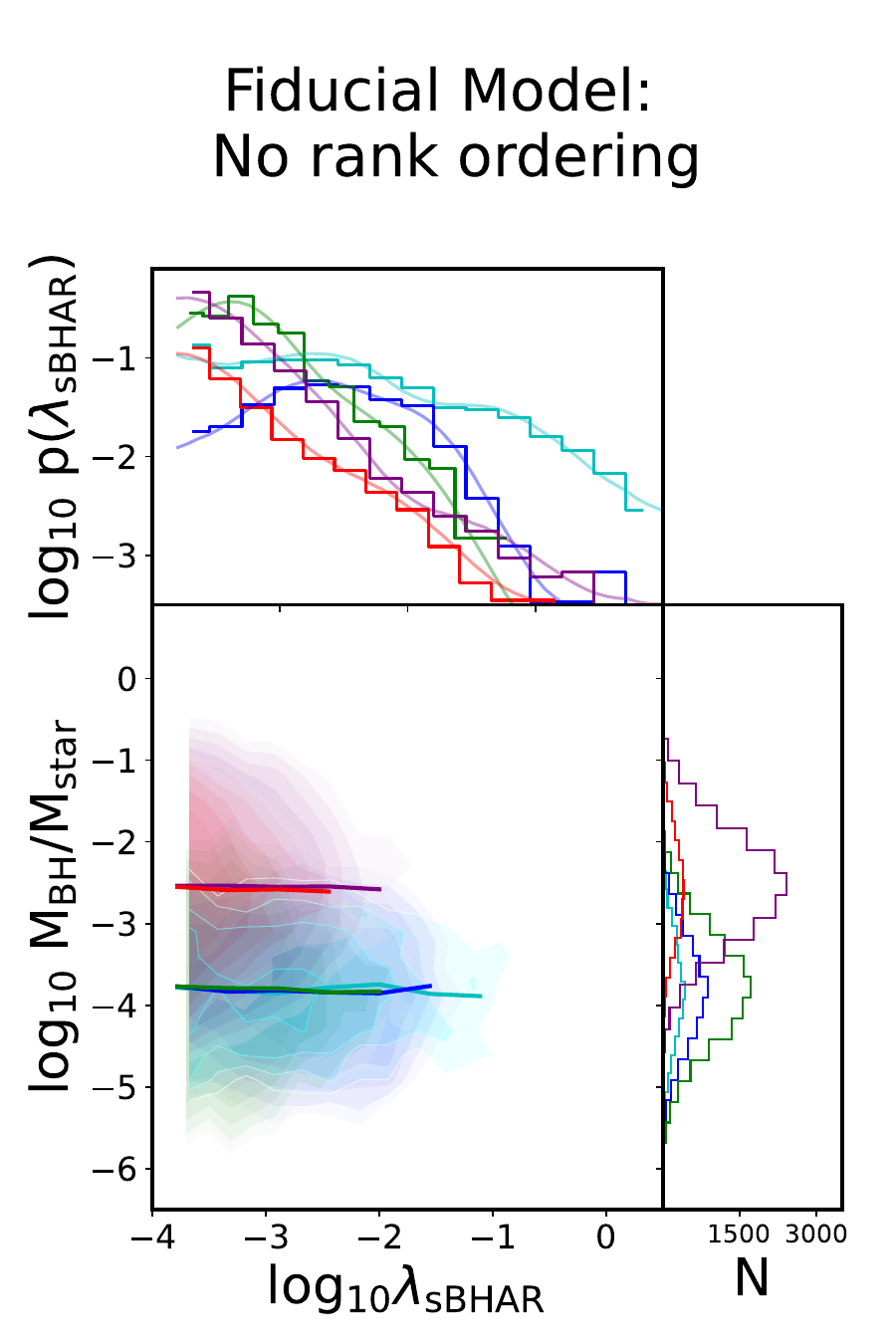}
\epsfxsize=5.6cm
\epsfbox{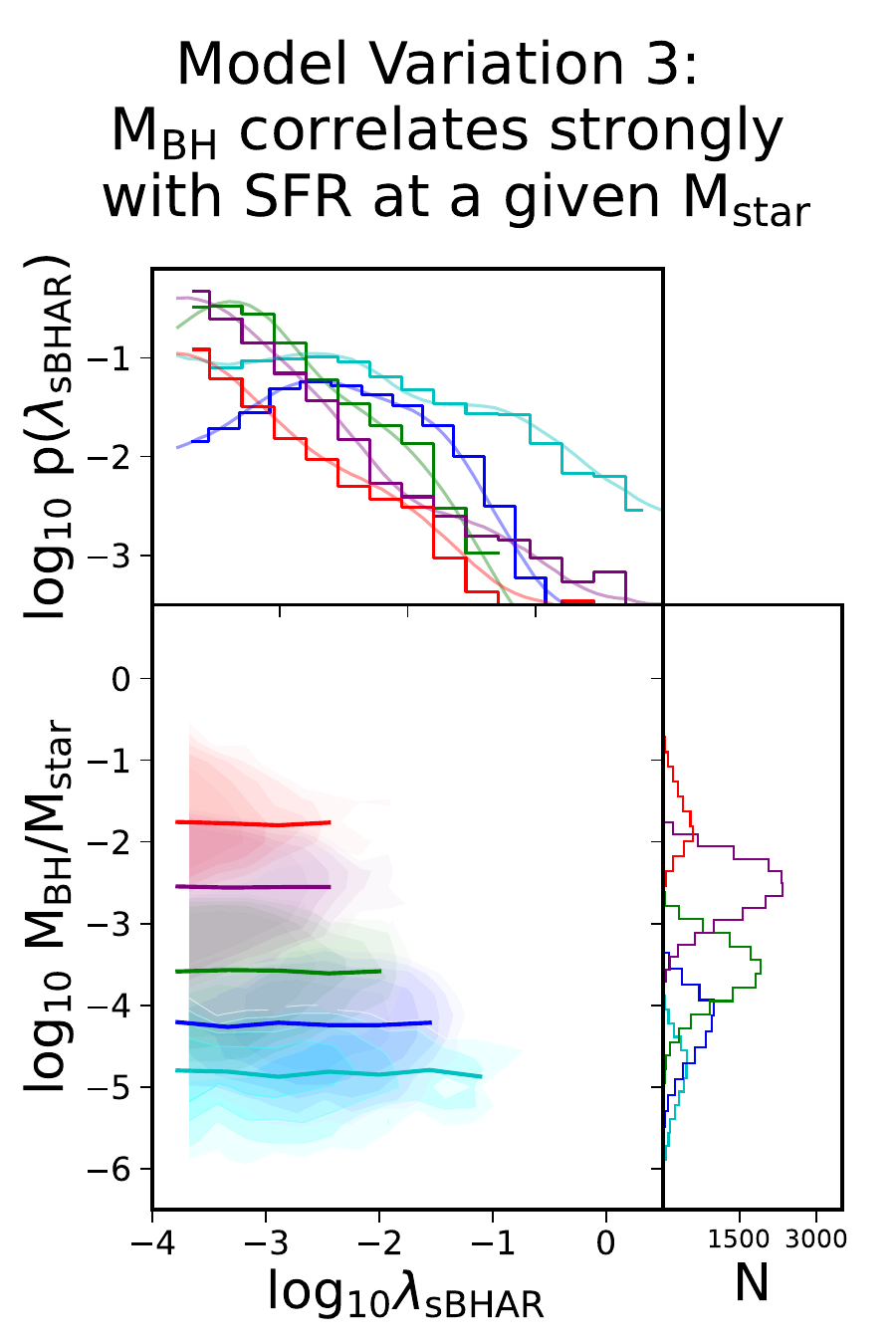}
\epsfxsize=5.6cm
\epsfbox{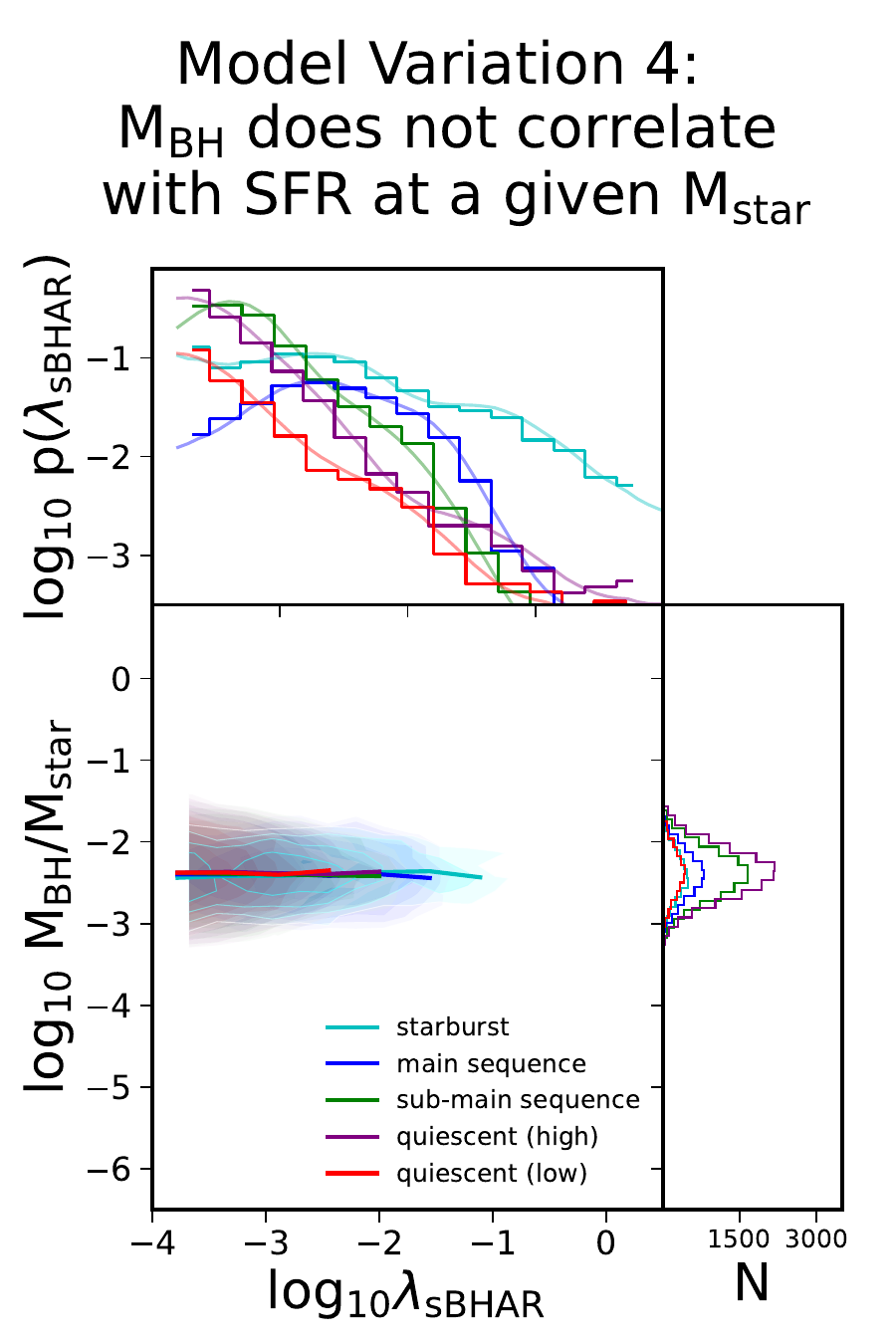}
\caption{The \mbh\---\mstar\ ratio as a function of \lambdasBHAR\ at $z = 0$ for the fiducial model and Model Variations 3 and 4, similar to Figure~\ref{fig:feddlambda} and Figure~\ref{fig:mvs}. The top panels show the derived probability distributions that result from our model (histograms), and the \citet{acg2019} probability distributions (smooth curves). These two curves for all models closely match one another by construction, regardless of the differences in how we assign $z = 0$ \mbh\ values in the model variations. The right panels show the histograms of \mbh\---\mstar\ ratio for our different SFR classifications, illustrating the different assignment of \mbh\ (relative to \mstar) for each model variation.}
\label{fig:mvs2}
\end{figure*}

Fig.~\ref{fig:mbhmstarmvs} and~\ref{fig:mvs2} summarize the differences between the fiducial model and these two additional model variations. Fig.~\ref{fig:mbhmstarmvs} shows the \mbh\---\mstar\ relations for each model at $z = 0$ where the colors indicate the correlation between SFR and our assignment of SMBH mass. In the leftmost panel SMBH masses are assigned to star-forming (blue) and quiescent (red) galaxies separately. In the center panel SMBH masses are assigned to galaxies based on their SFR, where the colors indicate the SFR bins used in this work (see Section~\ref{sec:method}). In the rightmost panel SMBH masses are assigned based on host galaxy \mstar\ alone, and thus the contours are shown by a single black color. The scatter in \mbh\ at a given \mstar\ is even more pronounced in Model Variation 3 compared to the fiducial model. The correlation between \mstar\ and \mbh\ in Model Variation 4 is much tighter than either of these two models and does not take into account observations of galaxies with lower \mbh\ for the same \mstar\ \citep{rv2015, tbh2016}. However, this model does reproduce similar trends that are seen in galaxy formation simulations where the relation is tight and evolves very little across redshift (e.g., IllustrisTNG, see \citealt{tbp2020}).

Figure~\ref{fig:mvs2} shows the relationship between the \mbh\---\mstar\ ratio and \lambdasBHAR\ for these three models (similar to Fig.~\ref{fig:feddlambda} and~\ref{fig:mvs}). Colors indicate our SFR classification bins and the top panels of each plot indicate that the probability distributions for \lambdasBHAR\ are reproduced for the different SFR bins in each of these models, regardless of the initial assignment of \mbh. However, the main panels and right-hand histograms illustrate the substantial differences in the range of \mbh\---\mstar\ ratios and how these relate to the galaxy SFRs in the fiducial model compared to these two model variations.

\begin{figure*}[!t]
\centering
\epsfxsize=15cm
\epsfbox{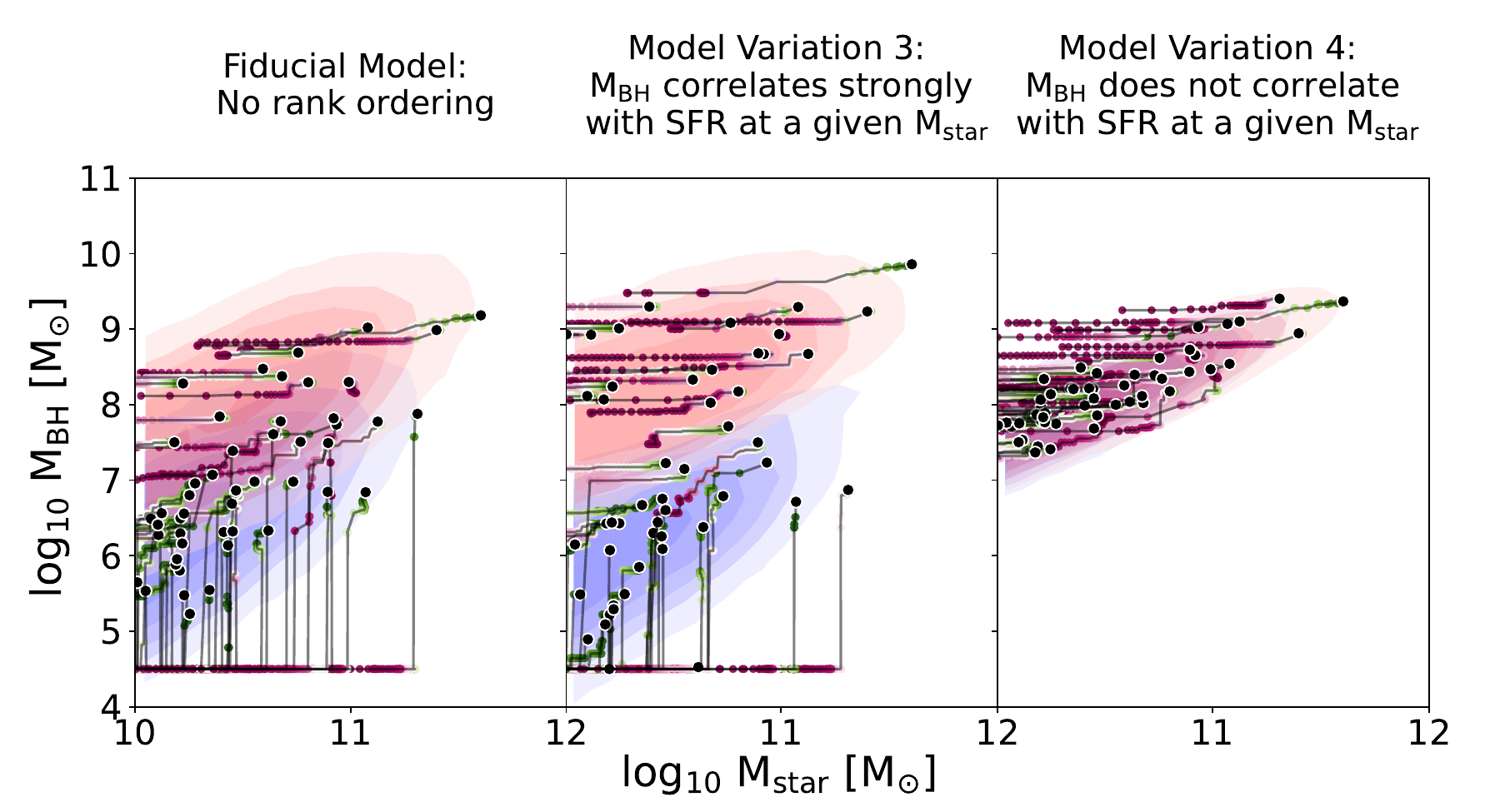}
\caption{Individual galaxy growth histories on the \mbh\---\mstar\ relation for our fiducial model (left panel) compared to Model Variation 3 (center panel) and Model Variation 4 (right panel), similar to Figure~\ref{fig:tracks} and ~\ref{fig:tracksmvs}. Model Variation 3 introduces a stronger correlation between \mbh, \mstar, and SFR at $z = 0$ compared to the fiducial model, but the same general pattern of flat tracks for the most massive $z = 0$ SMBHs and vertical growth tracks for the lower mass $z = 0$ SMBHs remains. Model Variation 4, which adopts a tight correlation between \mbh\ and \mstar\ with no dependence on SFR and minimal scatter leads to a much narrower range of growth trajectories, resulting in mostly flat, horizontal tracks.}
\label{fig:tracksmvs2}
\end{figure*}

\begin{figure*}
\centering
\epsfxsize=5cm
\epsfbox{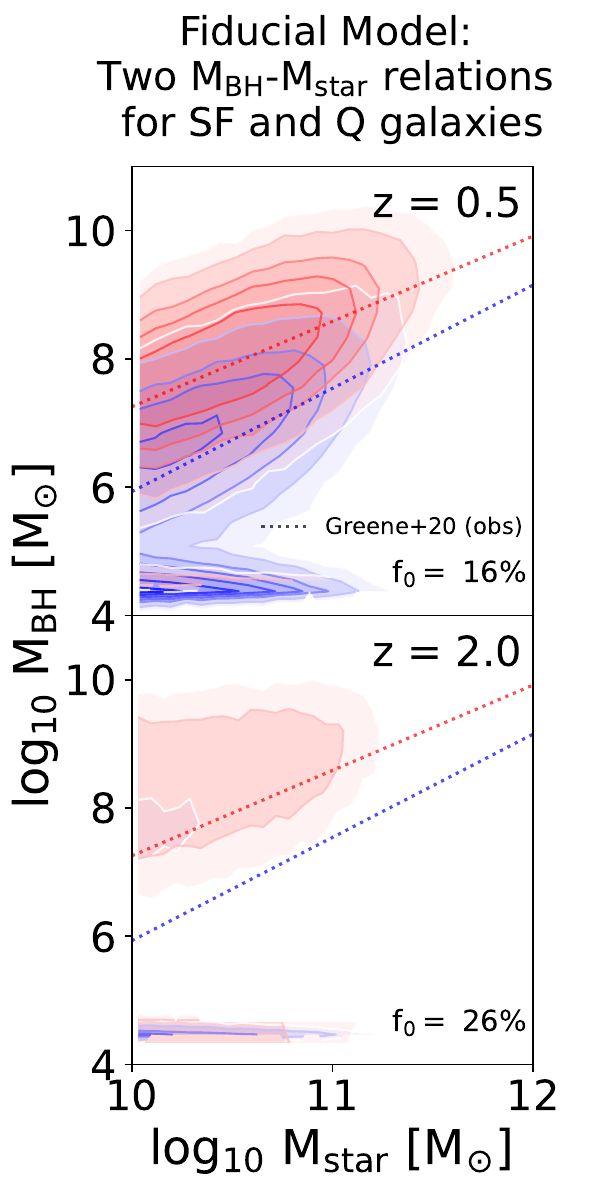}
\epsfxsize=5cm
\epsfbox{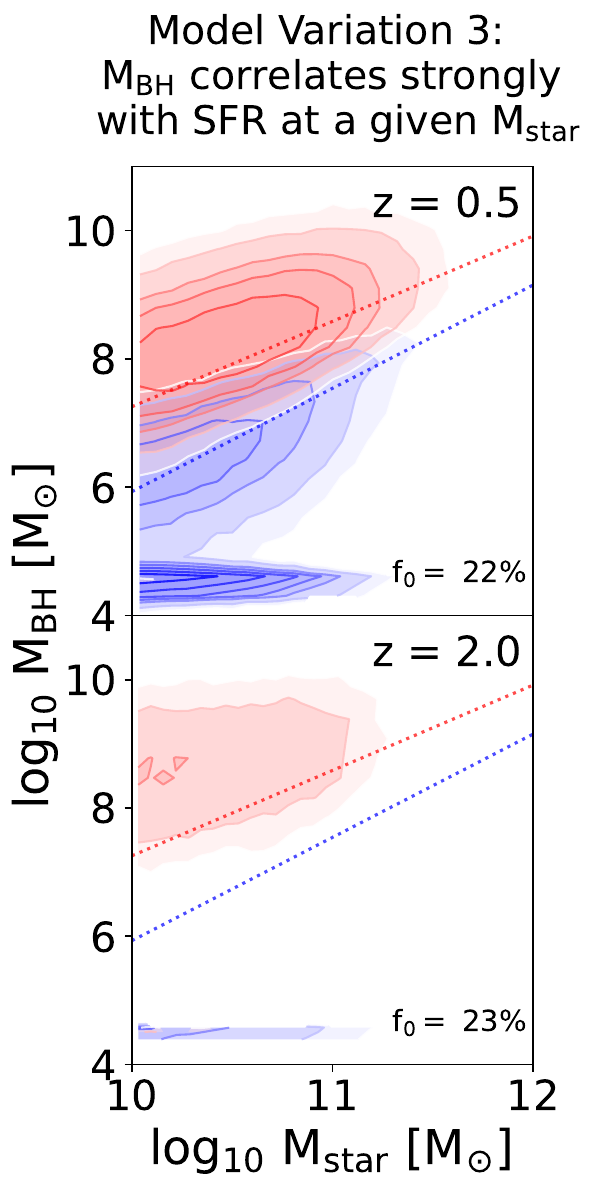}
\epsfxsize=5cm
\epsfbox{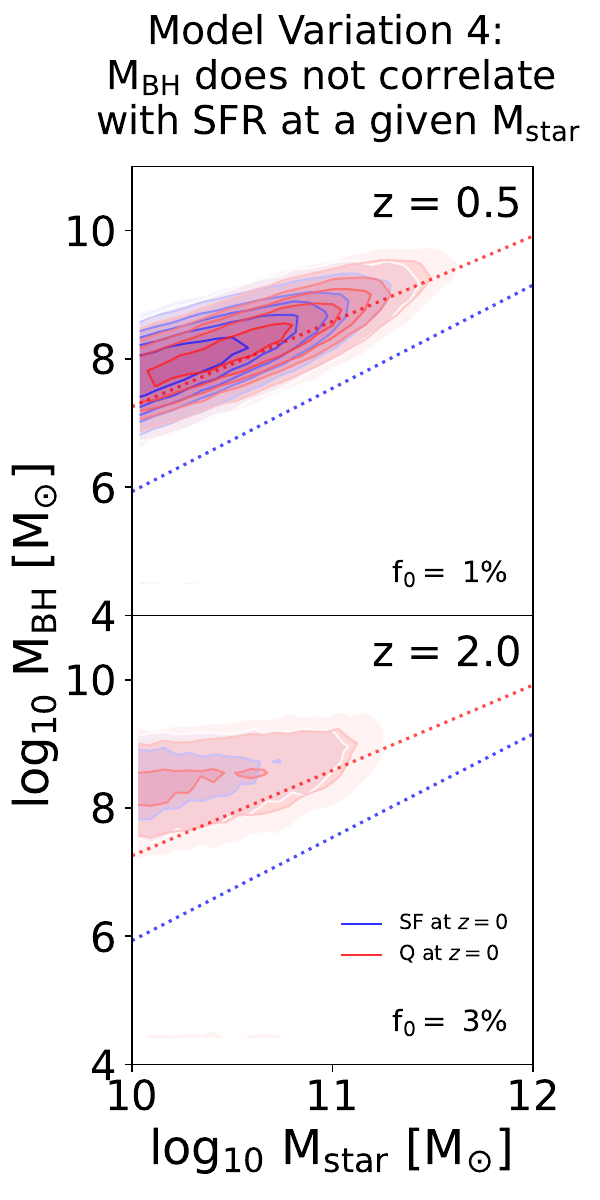}
\caption{The $z = 0.5$ and $z = 2$ \mbh\---\mstar\ relations for the fiducial model and Model Variations 3 and 4, similar to Figure~\ref{fig:evolvemvs}. The $z = 0$ \citet{gsh2020} relations are shown by the dotted lines for comparison in every panel. Similar to our findings comparing Model Variations 1 and 2, the differences in the method for assigning $z = 0$ \mbh\ distributions across the galaxy population result in similar relations at $z = 2$. However, for Model Variation 3 the distinction at $z = 2$ between galaxies that will be star-forming or quiescent at $z = 0$ is stronger, with the star-forming galaxies all either moving outside the limits of the plot or falling in the $f_0$ regime, while quiescence by $z = 0$ is strongly associated with a high \mbh\ at $z = 2$. Model Variation 4 exhibits significantly less scatter at $z = 2$ than the other models, a negligible $f_0$, and no distinction based on the $z = 0$ galaxy star formation properties. All models result in a flatter \mbh\---\mstar\ relation at $z = 2$.}
\label{fig:evolvemvs2}
\end{figure*}

Fig.~\ref{fig:tracksmvs2} shows evolutionary tracks for individual galaxies in each of these three models (similar to Fig.~\ref{fig:tracks} and~\ref{fig:tracksmvs}). Fig.~\ref{fig:evolvemvs2} shows the \mbh\---\mstar\ relations for star-forming (blue) and quiescent (red) galaxies at $z = 0.5$ (top panels) and $z = 2.0$ (bottom panels) for all three models (similar to Fig.~\ref{fig:evolvemvs}).

Model Variation 3 adopts the highest degree of correlation between SFR, \mbh, and \mstar. This model approximately correlates the SFR to the \mbh/\mstar\ ratio at $z = 0$, assigning overmassive SMBHs to progressively more quiescent galaxies. This model retains, and somewhat enhances, the diversity of the growth pathways of individual galaxies. Those galaxies with the highest \mbh\ at $z = 0$, and thus by construction the lowest SFRs, undergo little relative growth in \mbh\ between $z = 0-2$, resulting in flat growth tracks in the center panel of Fig.~\ref{fig:tracksmvs2}. In contrast, vertical tracks are common for $z = 0$ galaxies with higher SFRs. The connection between quiescence at $z = 0$ and an overmassive \mbh\ at $z = 2$ is even stronger for Model Variation 3, while it is exclusively $z = 0$ star-forming galaxies where we are able to account for the entirety of the $z = 0$ \mbh\ (the $f_0$ locus indicated in Fig.~\ref{fig:evolvemvs2}).

In the model with a single, tight relationship between \mbh\ and \mstar\ (Model Variation 4) the tracks are largely along the relation with a few galaxies with slightly lower SMBH masses undergoing periods of modest SMBH growth. All of the galaxies in this model host high SMBH masses at $z = 0$, assuming all galaxies follow the \citet{kh2013} relation which is very similar to the relation for quiescent galaxies that we use in our fiducial model (see Section~\ref{sec:mbh}). These galaxies have less diversity in their growth histories compared to models with larger scatter in \mbh\ at a given \mstar. Between $z = 0$ and $z = 0.5$, the \mbh\---\mstar\ relation is largely unchanged for this model. In stark contrast to the fiducial model and Model Variation 3, $f_0$ is negligible for Model Variation 4 at both $z = 0.5$ and $z = 2$, indicating that such SMBH masses cannot be accounted for by the observed distribution of SMBH accretion rates.

At $z = 2$ we find that all models (fiducial and Model Variations 1--4) produce very similar \mbh\---\mstar\ relations. Regardless of our assumptions for assigning accretion rates or \mbh\ at $z = 0$, we produce a flatter \mbh\---\mstar\ relation with higher normalization by $z = 2$. The population of galaxies with low \mbh\---\mstar\ ratios on the $z = 0$ relation are largely gone by $z = 2$. As was discussed in Section~\ref{sec:highMbh} this is due to high sBHARs for lower mass SMBHs that result in many galaxies with \mbh\ $= 0$ \msun\ at $z < 2$, or many galaxies whose \mstar\ evolves to $< 10^{10}$ M$_{\odot}$ by $z = 2$. Our models all indicate that the most massive SMBHs at $z = 0$ grew very little in \mbh\ since $z = 2$, producing a flatter \mbh\---\mstar\ relation at $z = 2$. This finding is robust despite very different assumptions in the assignment of $z = 0$ SMBH masses and allocation of accretion rates in our model variations.



\section{Discussion}
\label{sec:discussion}


This work builds empirically-motivated SMBH growth histories on top of galaxy growth histories in the UniverseMachine model framework. We use observational constraints on the $z = 0$ \mbh\---\mstar\ relations for early-type and late-type galaxies separately to establish the boundary conditions from which to build our growth histories. We then use observationally-derived probability distributions of SMBH accretion rates at $z = 0-2$ based on host galaxy star formation properties to assign SMBH growth rates to the $z = 0$ galaxy population. We follow this procedure iteratively to build SMBH growth histories back in time to $z = 2$ and thus link galaxy and SMBH assembly. 

We also alter a few key assumptions of our fiducial model and introduce four other model variations to explore alternative approaches to building SMBH growth histories. The first two model variations adjust how we assign accretion rates to galaxies by assuming overmassive SMBHs (those with high \mbh\---\mstar\ ratios) have the highest accretion rates. The other two adjust how we assign \mbh\ values to galaxies at $z = 0$, first by assuming maximal dependence on SFR and second by assuming no dependence on SFR. All model variations reproduce the observational constraints on sBHARs at all redshifts and the $z = 0$ \mbh\---\mstar\ relation by construction.

\subsection{Scatter in the \mbh\---\mstar\ relation implies a diversity of \mbh\ growth histories}


We first note the importance of taking into account the significant scatter in \mbh\ at a given \mstar\ at $z = 0$. Galaxy formation simulations often use canonical scaling relations between \mbh\ and some measure of \mstar\ \citep[usually M$_{\rm{bulge}}$,][]{hr2004, mm2013, kh2013} to calibrate and compare their results to determine the success of their model. Studies of SMBH growth in large-scale cosmological simulations often find that \mbh\ and \mstar\ growth track one another closely and produce a tight relation between these quantities \citep[e.g.,][]{shc2008, svg2015, scb2015, sdh2015, vdp2016}. 

In Section~\ref{sec:varrel} we present model variations that show different assumptions for the $z = 0$ \mbh\---\mstar\ relation. We find that the model variations that take into account the substantial scatter in the \mbh\---\mstar\ relation (the fiducial model and Model Variations 1, 2, and 3) produce significantly more diverse growth histories compared with Model Variation 4 that assumes a tight relation. When all SMBHs follow the same tight scaling relation, their growth histories trace similar paths. In our case, Model Variation 4 traces mostly flat, horizontal tracks, since all of the SMBHs are overmassive when comparing to other model variations. However, when substantial scatter is included, as seen in observations, the growth histories become more diverse. Galaxies with relatively overmassive SMBHs show growth histories that are largely flat on the \mbh\---\mstar\ relation, indicating more stellar mass growth compared to SMBH growth. Galaxies with undermassive SMBHs show growth histories that are largely vertical, indicating the rapid growth of SMBHs at late times.

Our results are consistent with the findings of \citet{zh2023} who perform a study of over 10,000 broad-line AGN at $z \leq 0.35$ from SDSS. They measure the current SMBH masses, accretion rates, stellar masses and SFRs for this sample and find that galaxies with overmassive SMBHs show more stellar mass growth compared to their SMBH mass growth, producing horizontal tracks on the \mbh\---\mstar\ relation. They find that undermassive SMBHs show the opposite trend with more SMBH growth than stellar mass growth resulting in more vertically aligned tracks. While \citet{zh2023} only study low redshift ($z\leq 0.35$) SMBHs with bright accretion disks that are preferentially found in star-forming galaxies, our results indicate that such trends persist for the full population at higher redshifts and remain true when considering the intermittent duty cycle of AGN throughout a galaxy's lifetime. Both of our studies strongly indicate there exists a degree of diversity of pathways for SMBH growth in galaxies that correlates with the scatter observed in \mbh\ at a given \mstar.

We caution that galaxy formation models that do not reproduce this scatter in the \mbh\---\mstar\ relation will underestimate the diversity of coevolutionary histories between their SMBH population and their host galaxies. Almost every SMBH model in the latest state-of-the-art simulations connects SMBH accretion to a feedback mechanism that quenches star formation. While these models have broadly reproduced a growing quiescent population via SMBH feedback, the exact physical mechanism(s) for this feedback, how it couples to galaxy evolutionary processes, and its link to accretion processes has not yet been determined. By taking the diversity of coevolutionary growth histories into account, galaxy formation models will take a step towards understanding how the interplay between these physical processes functions to obtain the observed star formation properties of galaxies across cosmic time.


\subsection{Strong and consistent evolution of the \mbh\---\mstar\ relation}
\label{sec:discevol}

A goal of this work is to understand the evolution of the \mbh\---\mstar\ relation from $z = 2$ to its current form at $z = 0$. While our results show different growth histories for the different assumptions in our model variations, by $z \sim 2$ we largely find the same relation for all versions of our model. As redshift increases, the normalization of the \mbh\---\mstar\ relation increases while the slope decreases resulting in a flatter relation at higher \mbh. This behavior is produced by the fact that the overall accretion rate density of the universe between $z = 0-2$ cannot fully account for the mass growth of the most massive SMBHs. This finding remains consistent across all our model variations, including Model Variations 1 and 2 where we preferentially bias most of the overall accretion rate density of the universe to take place in galaxies with overmassive SMBHs. Instead, these galaxies have horizontal tracks on the \mbh\---\mstar\ relation, indicating modest stellar mass growth without much fractional SMBH mass growth. 

In contrast, galaxies hosting lower mass SMBHs at $z = 0$ show vertical tracks in our fiducial model, largely disappearing off the \mbh\---\mstar\ relation at higher redshifts. This result is in agreement with studies showing that substantial SMBH growth can take place in star-forming, disky galaxies due to secular processes without any requirement for mergers to trigger accretion growth \citep{sls2013, ssl2017}. The combination of horizontal tracks with modest stellar mass growth for high mass SMBHs and vertical tracks for lower mass SMBHs results in a flatter high-redshift \mbh\---\mstar\ relation with a higher normalization. In the Model Variations where we bias SMBH growth to preferentially take place in overmassive SMBHs, undermassive SMBHs at $z = 0$ show horizontal tracks. However, since these galaxies with undermassive SMBHs are star-forming at $z = 0$, they tend to rapidly grow their stellar masses at all redshifts and the majority end up with \mstar\ $< 10^{10}$ M$_{\odot}$. Therefore, the area occupied by the undermassive SMBH population in the $z = 0$ \mbh\---\mstar\ relation largely depopulates with increasing redshift regardless of model variation.

These results indicate that there are populations of galaxies that grow their SMBHs early ($z \gg 2$) and other populations that assemble the bulk of their SMBH masses at later cosmic times ($z < 2$). In our work, we find that the galaxies in this latter category are able to host SMBHs with accretion rates that can easily account for the entirety of their $z = 0$ SMBH masses (refer to $f_0$ in Fig.~\ref{fig:evolution},~\ref{fig:forecast},~\ref{fig:evolvemvs}, and~\ref{fig:evolvemvs2}). This may indicate the need for galaxy formation models to adjust their seeding or growth prescriptions to allow for the gradual and continuous seeding of SMBHs at all redshifts to reflect the full diversity of \mbh\ growth histories. We find that this population consistently adds to the undermassive envelope of the \mbh\---\mstar\ relation and thus increases the scatter as galaxies and SMBHs assemble their masses from $z = 2$ to 0. The $z = 0$ \mbh\---\mstar\ relation is therefore assembled by SMBHs growing their mass at different times and gradually adding scatter to the $z = 0$ relation.

Recent work by \citet{zbv2023} presents an alternative empirical model of galaxy and SMBH growth, \textsc{Trinity}. Rather than modeling the growth histories of individual galaxies and their SMBHs, as in our model, \textsc{Trinity} traces the \emph{average} growth histories of SMBHs and galaxies for different halo mass bins. As such, \textsc{Trinity} follows the ensemble properties of SMBHs over cosmic time but is unable to trace the diversity in the individual growth trajectories revealed in our work. 

It is also worth noting that \textsc{Trinity} does not distinguish between star-forming and quiescent galaxies in their approach. Instead, average galaxy SFRs in halo mass bins are used to assign a single average BHAR to the galaxies in these bins. We have shown that the distinct scaling relations of star-forming and quiescent galaxies at $z = 0$ (and their scatter) result in diverse pathways for SMBH assembly, which will not be captured by the \textsc{Trinity} approach. Indeed, \textsc{Trinity} predicts relatively little evolution in the \mbh\---\mstar\ relation across the entire $z = 0$ to $z = 10$ range for galaxies with \mstar\ $\gtrsim3\times10^{10}$~\msun\ and a mild \emph{decrease} in normalization at lower stellar masses as redshift increases. In contrast, we find that the scaling relation steepens somewhat towards low redshift, although we caution that a single, average relation does not capture the diversity of growth pathways, the substantial scatter in the relation, and the ongoing seeding of SMBHs in star-forming galaxies toward later cosmic times that are critical features of our model. The greater diversity of \mbh\ at a given \mstar\ across cosmic time predicted by tracing individual growth histories, as we do in our model, may help alleviate some of the difficulties \textsc{Trinity} has in producing populations of overmassive SMBHs in intermediate-mass galaxies at $z\gtrsim5$ \citep[see][]{zbv2024} such as those found in recent JWST studies \citep[e.g.][also see discussion in Section~\ref{subsec:massiveSMBHs}]{msc2023,ggs2023,bgn2024}.



\subsection{The most massive SMBHs are established by $z = 2$}
\label{subsec:massiveSMBHs}

We find that the most massive SMBHs in the local universe are very difficult to grow at $z < 2$ given the SMBH growth rates found in \citet{acg2019}. Even when introducing significant bias and preferentially assigning the highest growth rates to the highest SMBH masses (see Model Variations 1 and 2 in Section~\ref{sec:highMbh}), we find that these SMBH growth histories show largely flat, horizontal evolutionary tracks on the \mbh\---\mstar\ relation. This indicates that the observed $z = 0$ scatter in \mbh\ across $\sim$5 orders of magnitude has important implications for determining the fractional SMBH mass growth across the galaxy population between $z = 0-2$. While accretion can easily account for the entirety of SMBH mass assembly in a large fraction of galaxies with undermassive SMBHs at $z = 0$, this is not the case for overmassive SMBHs. Instead, we find that the highest mass SMBHs found in the local universe must already be in place before $z = 2$.

Similar conclusions are reached by \citet{gag2024}, who present updated measurements of sBHAR probability distributions as a function of \mstar\ in massive galaxy populations using data from the $\sim9$~deg$^2$ \textit{Chandra} Deep Wide-field Survey \citep{mhc2020}. They use these probability distributions to calculate average accretion rates of individual SMBHs over time and thus trace their mass assembly since $z \sim 4$. They find that the most massive SMBHs at $z = 0$ must have grown relatively little of their final mass over this time period. In contrast, they use observed average SFRs to find that a significant fraction of the stellar mass in these galaxies assembled during the same period, thus predicting the existence of overmassive SMBHs in moderate-mass galaxies at $z\gtrsim4$. Our modeling extends on this approach, using the UniverseMachine to self-consistently track the galaxy population over time, assigning varying levels of SMBH growth based on the sBHAR probability distributions, and revealing the diverse pathways of \mbh\ assembly.

Our findings align with recent results obtained with JWST \citep{gma2023, rpm2023} that find not only high space densities of AGN in the early ($z > 5$) universe \citep{hzn2023, mnb2024, glg2024} but also that the SMBHs powering these AGN are typically overmassive relative to their host galaxy \mstar\ \citep{msc2023,slr2024,yes2024}. Our empirical model predicts the existence of such overmassive SMBHs at $z > 2$, suggesting that they are the progenitors of the highest mass SMBHs in the local universe. We find that these SMBHs are hosted in galaxies that follow horizontal tracks across the \mbh\---\mstar\ parameter space from $z\sim2$ to $0$, growing relatively little in \mbh\ but continuing to assemble their \mstar. The existence of these overmassive SMBHs out to even earlier cosmic times suggests that SMBH assembly continues to follow the diverse pathways we highlight in this work, although it remains unclear whether these high SMBH masses are assembled via rapid accretion at yet higher redshifts or whether SMBHs can form directly with such extreme masses. \newline


\subsection{Caveats}
\label{sec:caveats}

Throughout this work we have emphasized the importance of treating star-forming and quiescent galaxies differently, particularly in assigning SMBH masses at $z = 0$ according to different observed relations for these two populations \citep{tbh2016, tbw2017, gsh2020}. Ideally, we would be able to reliably predict \mbh\---\mstar\ relations that would be observed for star-forming and quiescent galaxies at higher redshifts, if we could obtain reliable SMBH masses. However, we find that the UniverseMachine produces star formation histories that limit our ability to reliably identify the onset of quiescence for individual galaxies.  We find that $\sim$ 97\% ($\sim$ 64\%) of UniverseMachine galaxies exhibit $>2$ ($>10$) transitions between star-forming and quiescent populations between $z = 0-2$. Several studies also find that individual star formation histories in the UniverseMachine are significantly more bursty than physically-motivated galaxy formation models due to the statistical nature of their model \citep{itg2020,ahb2023}.

This results in a substantial amount of mixing of the quiescent and star-forming populations when building our growth histories backwards in time. Fig.~\ref{fig:evolution} shows that by $z = 0.5$ the two populations are largely indistinguishable on the \mbh\---\mstar\ relation. We believe this is a result of the statistical nature of the UniverseMachine and not a physically meaningful result of the model. For this reason we primarily show whether galaxies are star-forming or quiescent based on their SFRs at $z = 0$ (see Fig.~\ref{fig:forecast}). We note that, despite the mixing between populations, the ensemble population of simulated UniverseMachine galaxies are, by construction, made to reproduce the stellar mass functions and quenched fractions across redshift. Thus, while UniverseMachine may not fully capture the evolutionary paths of individual galaxies and their properties across time, it does faithfully capture the bulk average of this evolution. 

Despite these limitations, we find strong differences between \textit{individual} growth histories for the population of galaxies with overmassive SMBHs compared to those with undermassive SMBHs. This is true even when making different assumptions in the assignment of SMBH growth rates to individual galaxies in our model variations, indicating our results are robust to these changes.

We also note that \citet{acg2019} use only one \mstar\ bin, choosing instead to bin their data more finely using SFR. \citet{acg2018} split the same data between two SFR bins and several \mstar\ bins, probed to lower \mstar\ at lower redshifts, and found differing stellar-mass dependencies in the AGN incidence as a function of redshift. A comparison study building a similar model but incorporating such stellar-mass dependencies would be useful to compare with and extend our results. 

We also note that \citet{acg2019} use X-ray detections for quantifying the sBHAR. While X-ray data provide a clear indication of accretion in some systems, it may not reliably track the underlying accretion rate across the entire AGN population. X-ray selection also misses the most heavily obscured systems. We have adopted a simple correction that assumes a comparably sized population of heavily obscured sources with the same distribution of sBHARs \citep[see][]{lga2024} and that such sources occupy galaxies in the same manner as their unobscured counterparts.

Although other AGN selection methods may be even more susceptible to their own biases, future work using multiwavelength observations (including X-ray data) to constrain the probability distributions of sBHAR would reveal whether X-ray detections are particularly biased and in what systems they might be biased towards. The data from \citet{acg2019} also leave out the small fraction of quasars from their sample since reliable SFRs were difficult to determine for these sources. \citet[see appendix C therein]{acg2018} demonstrate that the exclusion of such sources does not drastically alter the shape of the measured accretion probability distributions, but it may have a more substantial impact on the inferred SMBH mass growth if a substantial fraction occurs in these high-luminosity phases, particularly toward higher redshifts. However, \citet{gag2024} include quasar-dominated sources (adopting both AGN and galaxy components in their SED fitting) and reach similar conclusions to our work, indicating that any additional \mbh\ growth is comparatively rare and short-lived quasar phases since $z = 4$ are not sufficient to account for the most massive SMBH at $z = 0$.

\section{Conclusions}
\label{sec:conclusions}

The primary motivation of our work is to illuminate how populations of central SMBHs grow alongside their host galaxies. We take key observational constraints on SMBH accretion rates as a function of star formation rate at $z = 0-2$ and the \mbh\---\mstar\ relation at $z = 0$ for galaxies with dynamically-measured SMBH masses. Using these data, we assign SMBH accretion rates at every snapshot of the UniverseMachine empirical model of galaxy formation in post-processing. With the growth rates assigned, the $z = 0$ \mbh\---\mstar\ relation for star-forming and quiescent galaxies serve as the boundary conditions from which we build SMBH growth histories from $z = 0$ backwards in time to $z = 2$. 

The SMBH accretion rate constraints are provided by \citet{acg2019} in the form of probability distributions of specific black hole accretion rates (\lambdasBHAR\ $\propto L_\mathrm{bol}/M_\mathrm{star}$) as a function of distance from the star-forming main sequence. We use the \mbh\---\mstar\ relations for star-forming and quiescent galaxies at $z = 0$ from \citet{gsh2020} where quiescent galaxies have more massive SMBHs than star-forming galaxies at a given stellar mass.

We find it particularly useful to explore various assumptions of our model via 4 model variations. In the first two, we preferentially bias accretion to occur in galaxies with overmassive SMBHs to determine if our model can account for their mass assembly since $z = 2$. In the other two, we alter the $z = 0$ correlation between \mbh, \mstar, and SFR that serve as the boundary conditions for our model to determine how sensitive our results are to this assumption. All model variations reproduce the observed sBHAR probability distributions from \citet{acg2019} by construction.

We summarize our results as follows:

\begin{itemize}
  \item There is not one answer to the question of whether SMBHs grow before galaxies assemble their stellar mass or vice versa. Instead, we find that galaxies with overmassive SMBHs must have assembled most of their \mbh\ at $z > 2$, while galaxies with relatively undermassive SMBHs at low redshift have accumulated their masses gradually across $z = 0-2$. 
  \item The observed scatter in the \mbh\---\mstar\ relation at $z = 0$ spans $\sim4-5$ orders of magnitude and indicates a diversity of coevolutionary pathways for SMBHs and their host galaxy stellar mass. At higher redshifts, the scatter in our model remains substantial. We find that assuming instead that all galaxies follow a single, relatively tight \mbh\---\mstar\ relation at $z = 0$ greatly reduces the diversity of possible SMBH-galaxy coevolutionary growth histories.
  \item The most massive SMBHs observed today are extremely difficult to grow since $z = 2$ using current constraints of SMBH accretion rates. No matter how strongly we bias SMBH growth to occur in quiescent galaxies with substantially overmassive SMBHs ($> 10^{8}$ M$_{\odot}$), these SMBHs grow very little of their total mass since $z = 2$. This indicates that overmassive SMBHs grew very rapidly in the early universe before cosmic noon, a finding which is consistent with recent JWST discoveries of high space densities of AGN at $z\gtrsim5$, many of which may already have overmassive SMBHs relative to their hosts.
  \item We find that the \mbh\---\mstar\ relation evolves to have higher normalization and shallower slope at higher redshift for all model variations presented here. The most massive SMBHs display relatively little fractional \mbh\ growth since $z = 2$. This results in the $z = 2$ \mbh\---\mstar\ relation for \mstar\ $>10^{10}$ \msun\ being almost entirely made up of galaxies that become quiescent by $z = 0$. In contrast, undermassive SMBHs at $z = 0$ rapidly accumulate the bulk of their masses at late times, suggesting that lower mass SMBHs are seeded gradually throughout cosmic time.
\end{itemize}

The model presented here is entirely empirical and relies on very few physical assumptions (i.e., the link between observed X-ray luminosity and SMBH mass accretion rate). Our empirical approach is computationally inexpensive and has the advantage of allowing a flexible exploration of several model variations that are useful for testing assumptions and understanding the implications of the observational data with few physical priors. Our work has shown this approach to be a provide a powerful tool for exploring SMBH growth histories and serves as a complimentary approach alongside physically-motivated modeling of SMBH mass assembly.

\begin{acknowledgements}

We thank Rachel Somerville, Greg Bryan, and Peter Behroozi for useful discussions. We also acknowledge the UniverseMachine developers for their public datasets that made this work possible. J.A. acknowledges support from a UKRI Future Leaders Fellowship (grant code: MR/T020989/1). For the purpose of open access, the authors have applied a Creative Commons Attribution (CC BY) license to any Author Accepted Manuscript version arising from this submission.

\end{acknowledgements}

\bibliography{paper}{}
\bibliographystyle{aasjournal}

\end{document}